\pdfoutput=1
\documentclass{JINST}

\usepackage{url} 
\usepackage{tabularx}
\usepackage{lineno}

\title{First operation and drift field performance
of a large area double phase LAr  Electron Multiplier Time Projection Chamber with an 
immersed Greinacher high-voltage multiplier} 

\author{A.~Badertscher$^a$, A.~Curioni$^a$, U.~Degunda$^a$, L.~Epprecht$^a$, A.~Gendotti$^a$,
S.~Horikawa$^a$, L.~Knecht$^a$, D.~Lussi$^a$, A.~Marchionni$^a$, G.~Natterer$^a$, K.~Nguyen$^a$, F.~Resnati$^a$, 
A.~Rubbia$^a$\thanks{Corresponding author.}, and T.~Viant$^a$~\\
\llap{$^a$}ETH Zurich, Institute for Particle Physics,\\
  CH-8093 Z\"{u}rich, Switzerland\\
  E-mail: \email{andre.rubbia@cern.ch}}

\abstract{
We have operated a liquid-argon large-electron-multiplier time-projection chamber (LAr LEM-TPC) with
a large active area of 76 $\times$ 40 cm$^2$ and a drift length of 60~cm.
This setup represents the  largest chamber ever achieved with this novel detector concept. 
The chamber
is equipped with an immersed built-in cryogenic Greinacher multi-stage
high-voltage (HV) multiplier, which, when subjected to an external AC HV of $\sim$1~kV$_{\mathrm{pp}}$,
statically charges up to a voltage a factor of $\sim$30 higher inside the LAr vessel, creating
a uniform drift field of $\sim$0.5~kV/cm over the full drift length. 
This large LAr LEM-TPC was brought 
into successful operation in the double-phase (liquid-vapor) operation mode and tested during a period of 
$\sim$1~month, recording impressive three-dimensional images of very high-quality  from 
cosmic particles traversing or interacting in the sensitive volume.
The double phase readout and 
HV systems achieved stable operation in cryogenic conditions demonstrating their good characteristics, 
which particularly suit applications for next-generation giant-scale LAr-TPCs. 
}


\begin{document}

\section{Introduction}
\label{sec:1}
%

The design concept of the novel Liquid Argon Large Electron Multiplier Time Projection Chamber (LAr LEM-TPC) and its promising performances have 
been demonstrated experimentally with a detector prototype having an active area of 10 $\times$ 10 cm$^2$ 
\cite{Badertscher:2010zg, Badertscher:2009av}. 
This new type of LAr-TPC, operating in the double-phase (liquid-vapor) mode in pure argon,
is characterized by 
its charge
amplifying stage (LEM) 
and its projective anode charge readout system. 
Situated in the vapor phase on top of the active LAr volume, these latter provide an adjustable charge gain 
and two independent readout views, presently each with a pitch of 3~mm\footnote{Thanks to the amplification stage
which yields enough charge to be shared among several readout electrodes, smaller pitch sizes can be considered.}. 
With $T_0$ time information provided by the LAr scintillation, they 
provide a real-time three-dimensional (3D) track imaging 
with $dE/dx$ information 
and the detector acts as a high resolution tracking-calorimeter. 
A very high signal-to-noise ratio  can be reached in the LAr LEM-TPC thanks to the gaining stage.
This significantly  improves the 
event reconstruction quality 
with a lower energy deposition threshold and a better resolution per volumetric pixel (voxel) 
compared to a conventional single-phase LAr-TPC 
\cite{Badertscher:2009av}, such as e.g. ICARUS~\cite{Amerio:2004ze},
ArgoNeut~\cite{Anderson:2011ce} or J-PARC T32~\cite{Araoka:2011pw}. 
In addition the charge amplification 
compensates for potential loss of signal-to-noise due to the charge diffusion and
attachment to electronegative impurities diluted in LAr, 
which both become more important as the drift length increases. 
It is therefore considered to be a promising technology for the next-generation giant-scale underground detectors for neutrino physics and proton decay searches (GLACIER)~\cite{Rubbia:2004tz,Rubbia:2009md,Badertscher:2010sy} 
and for direct Dark Matter searches with imaging \cite{Rubbia:2005ge}. 

In this paper, we report on the first operation of a LAr LEM-TPC prototype having an active area of 76 $\times$ 40 cm$^2$ and a drift length of 60 cm, 
which is the largest chamber ever realized with this novel detector concept. 
The detector was tested successfully in the double-phase operation mode in pure argon, recording events of cosmic tracks traversing or interacting
in the active volume. 
The  analysis of its tracking performance which requires full 3D reconstruction of a large number of recorded tracks is reported elsewhere~\cite{Devis:electronics}, 
and the focus of this paper is put on the immersed high-voltage (HV) system for the drift field and cold operation. 

The HV system is based on the technique of a cryogenic Greinacher \cite{Greinacher} (also known as Cockcroft-Walton \cite{CW}) HV multiplier 
immersed in LAr directly inside the detector vessel. 
It is an innovative technique 
to generate a HV for creating a drift electric field in a LAr-TPC, with some advantages compared to other methods
used up to now whereby the very HV is externally generated and brought into the detector via a HV feedthrough
(for a comparison see Ref.~\cite{Horikawa:2010bv}).
This technique particularly suits a large- to giant-scale detectors, where a very HV of $\sim$100 kV to 
$\sim$1 MV 
would be needed for a drift field of $\sim$1 kV/cm over a length of 
$\sim$1 m to $\sim$10 m.

The main advantages of the Greinacher HV multiplier are:
(1) all the HV parts are immersed in LAr which has a large dielectric strength ($\sim$1 MV/cm), (2) thus feedthroughs for very HV are not needed, (3) the circuit itself can provide all the stages needed to create a uniform field, thus the system has no resistive load, (4) thus the power dissipation is virtually zero and (5) this allows a low frequency (e.g. 50 Hz) of the AC input signal which is fully outside of the bandwidth of the charge amplifiers used for this type of detectors. 
One main disadvantage is the difficultly of access in case of a failure. The potential
impact of this access on the pure liquid argon volume can be large if the active components are directly 
immersed in it\footnote{It is still possible to imagine that the Greinacher circuit is located
in a separately evacuable volume in contact with the main pure liquid argon volume during normal operation,
but which would be disconnected and separately emptied during maintenance, with minimal
impact on the main liquid argon volume, thereby still avoiding the need for a HV feedthrough.}.

To study in detail the functionality of such a system under practical conditions incorporated in a LAr-TPC, we built a 30-stage Greinacher HV multiplier integrated to the large-area LAr LEM-TPC prototype. 
We carried out the test experiment at CERN from October to December 2011 using cryogenic LAr apparatus of the ArDM Project (CERN RE18) 
\cite{Marchionni:2010fi},
which is an experiment for a direct Dark Matter search exploiting a 1-ton-scale double-phase argon TPC~\cite{Rubbia:2005ge}. 

In Section~\ref{sec:2} we describe the experimental setup, i.e. the LAr LEM-TPC, the cryogenic apparatus and the HV system. 
Results of a series of measurements performed in air at room temperature and in LAr for characterizing the Greinacher HV multiplier are presented in Section~\ref{sec:3}. 
In Section~\ref{sec:4} we report on the test experiment, where the entire LAr LEM-TPC with the HV system was tested successfully 
in the double-phase operation mode in pure argon. 
Finally we discuss and conclude on the results in Section~\ref{sec:5}.

\section{Experimental setup}
\label{sec:2}
\subsection{Design of the large-area LAr LEM-TPC}
\label{sec:2.1}
\subsubsection{LAr LEM-TPC design overview}
\label{sec:2.1.1}

\begin{figure}[hbtp]
\begin{center}
\includegraphics[height=20pc]{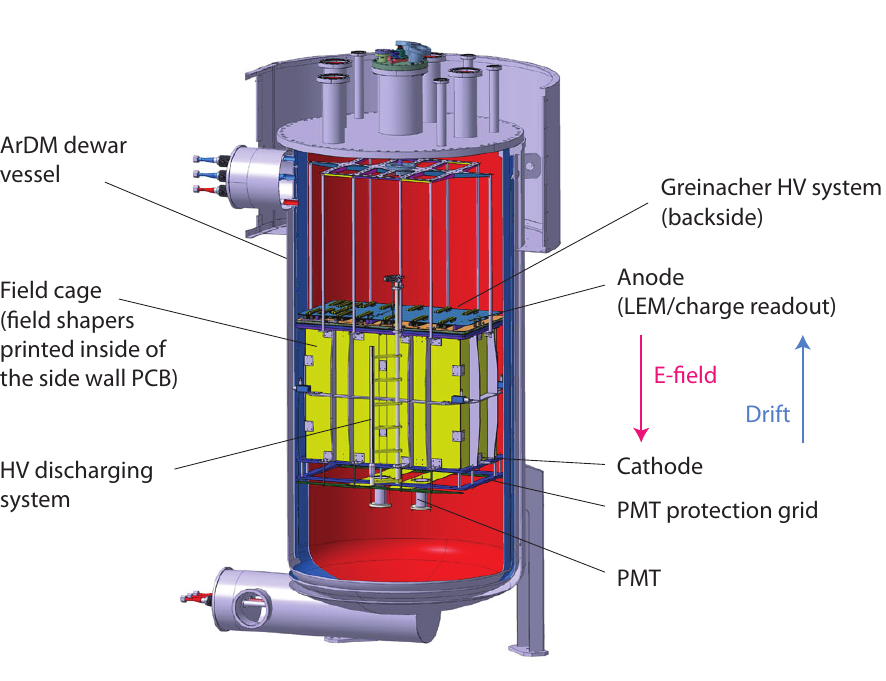}
\caption{3D CAD model of the LAr LEM-TPC prototype installed in the ArDM LAr vessel. The field cage has dimensions 76 cm (L) $\times$ 40 cm (W) $\times$ 60 cm (H). The LEM charge readout system is mounted on top of the field cage. Negative HVs linearly decreasing from bottom to top are generated by the built-in Greinacher HV multiplier and distributed to 31 field shapers formed by etching on the inner surface of the side wall PCBs. A vertical drift electric field thus is created uniformly over the full drift length of 60 cm.}
\label{fig:fig1}
\end{center}
\end{figure}

Figure \ref{fig:fig1} 
illustrates 
the structure of the LAr LEM-TPC hanging inside the ArDM detector vessel. 
The field cage, which defines the active LAr target and drift volume, is constructed with four PCB side walls in a shape of rectangular box with dimensions 
76 cm (L) $\times$ 40 cm (W) $\times$ 60 cm (H).
On the inner surface of each PCB 31 horizontal field shaping strips, which hereafter are called field shapers (FS), 
are formed by etching at a constant pitch of 20 mm.
The top (FS0) and the bottom (FS30) field shapers have a width of 9 mm while the ones in between have 18 mm, with a gap of 2~mm between neighboring ones. 
Negative high voltages linearly decreasing from bottom to top are distributed to the field shapers so that a vertical drift electric field is created uniformly over the full 
drift length of 
60 cm.

On top of the field cage two horizontal extraction grids are mounted with a gap of 10 mm in between. 
The grid is a 0.15-mm-thick 
stainless-steel mesh, where square holes with a size  2.85 $\times$ 2.85 mm$^2$ are etched at a pitch of 3 mm all over the active area. 
The lower grid is positioned at the top face of the field cage.
For the double-phase operation mode the LAr surface is adjusted at the middle of the two grids. 
The liquid level at each of the four corners can be monitored with a precision of $\sim$0.5 mm with the aid of four capacitive level meters. 
Ionization electrons produced by ionizing particles are drifted upwards to the liquid surface. 
These electrons are extracted across the liquid-vapor interface into the gas argon (GAr) phase 
with the aid of 
a strong extraction field of typically 3--4 kV/cm between the two grids. 
They are then collected by the charge readout 
system 
incorporating the Large Electron Multiplier (LEM), 
which is described in Section~\ref{sec:2.1.2}.
The bottom face of the field cage is covered by the cathode grid 
that is a stainless-steel mesh of the same type as used for the extraction grids.
The top (FS0) and the bottom (FS30) field shapers are electrically coupled to the lower extraction grid and respectively, to the cathode. 

\begin{figure}[hbtp]
\begin{center}
\includegraphics[height=17pc]{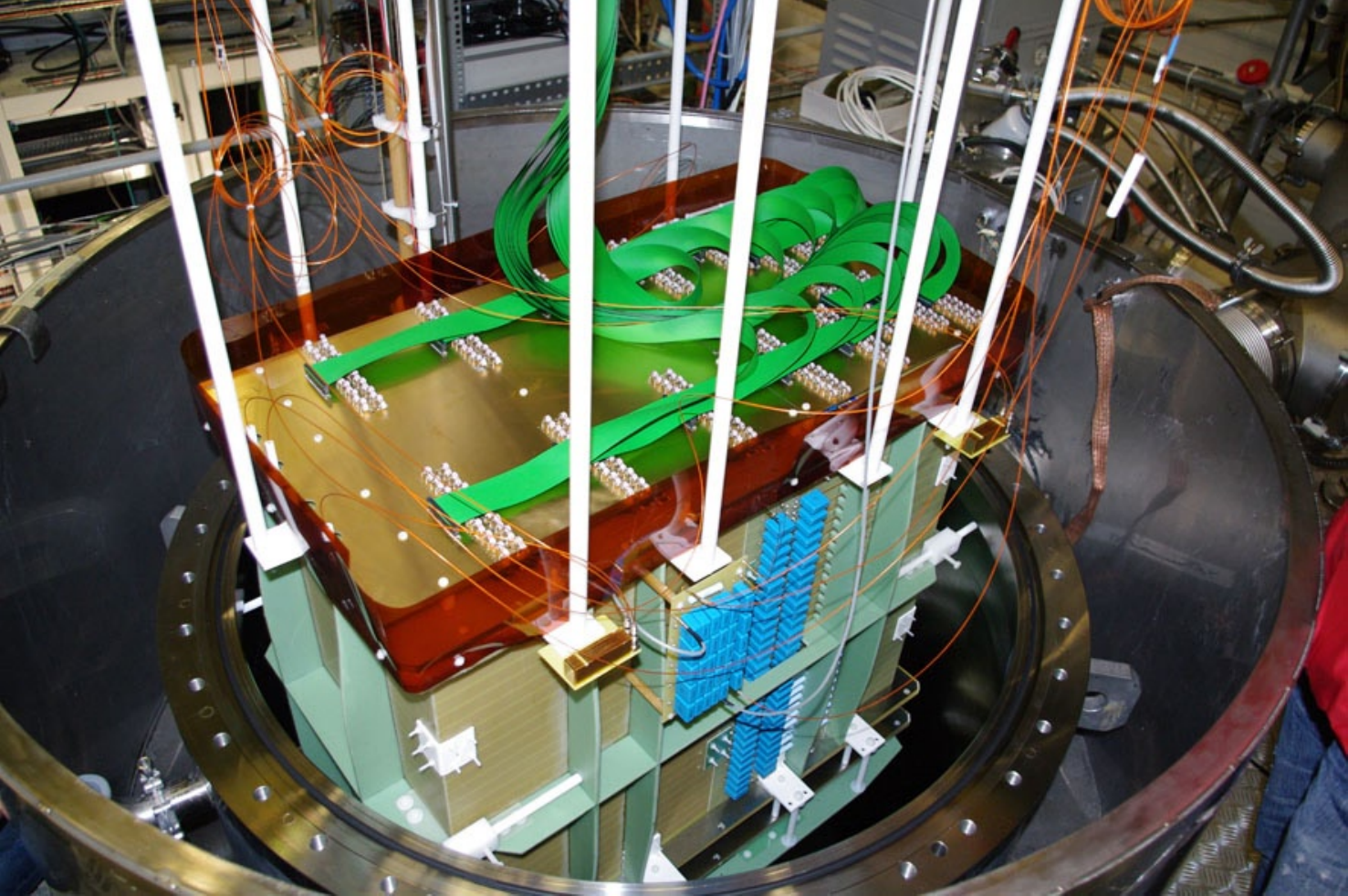}
\caption{Picture of the LAr LEM-TPC with readout sandwich
and drift field cage, which are about to be inserted into the ArDM vessel. 
The stack of blue capacitors on the front side wall shows the Greinacher HV multiplier.}
\label{fig:fig2}
\end{center}
\end{figure}

The HV for creating the drift field is generated using a built-in 30-stage Greinacher HV multiplier, which is described in detail in Section~\ref{sec:2.3}. 
The circuit is integrated in the design of one of the side-wall PCBs and the components are mounted directly on the outer surface of the PCB, 
as can be seen in Figure~\ref{fig:fig2}. 
As already mentioned the generator itself avoids the use of a voltage divider, since each multiplying stage provides
a characteristic DC voltage. The various multiplying stages generate a monotonously increasing potential to 
supply the electrodes surrounding the drift volume. In our setup, 
the DC output of each of the 30 Greinacher stages is connected to each field shaper via a wire through the PCB. 

Under the field cage two photomultiplier tubes (PMTs) are installed to detect scintillation light produced by charged particles crossing the LAr target (primary scintillation, S1), as well as the secondary scintillation light (S2) produced via electro-luminescence by the electrons extracted to the 
GAr phase 
\cite{Monteiro:2008zz}.
We employed Hamamatsu R11065 cryogenic 3'' PMTs having a high quantum efficiency of $\sim$30\%, specifically designed for the use in LAr 
\cite{Acciarri:2011qx}.
The quartz window of the PMT was coated with a wavelength shifter, 1,1,4,4-Tetraphenyl-1,3-butadiene (TPB) \cite{Boccone:2009kk}, 
in a Paraloid\texttrademark\, B-72 polymer matrix to detect 
deep ultra violet photons of the argon scintillation (peaked around 128~nm) . 
The PMT signal is used primarily for triggering and for determination of the event $T_0$ absolute time. 
To protect PMTs from possible discharges from the parts at HV, another mesh electrode, which is grounded to the detector vessel,  
is inserted between the cathode grid and the PMTs. 
The distance between the cathode and the protection grid was 9~cm and that between the protection grid and the PMTs 
was 1~cm.

\subsubsection{LEM charge readout system}
\label{sec:2.1.2}

\begin{figure}[hbtp]
\begin{center}
\includegraphics[width=\columnwidth]{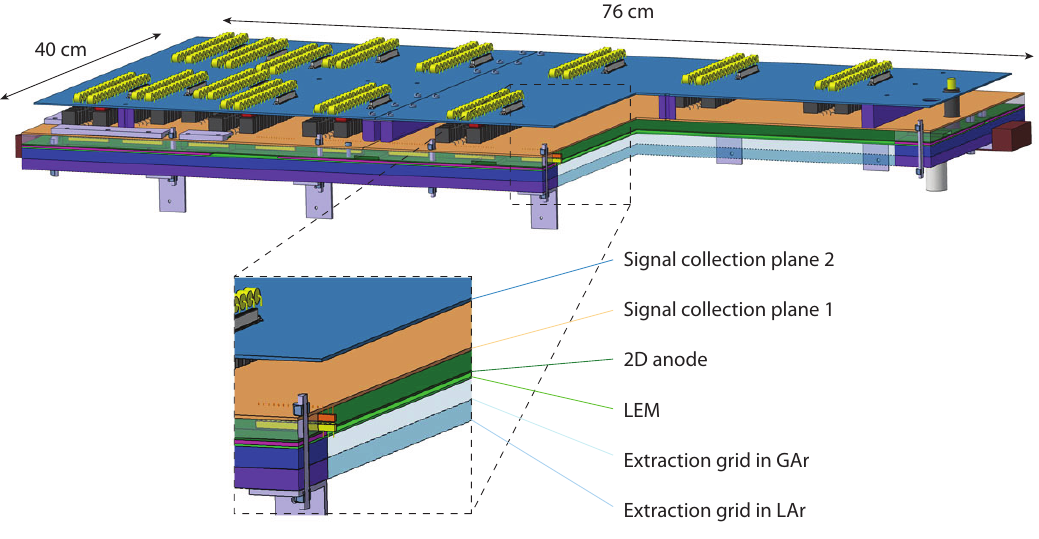}
\caption{Structure of the charge readout system incorporating a LEM (Large Electron Multiplier) and a 2D anode. Right-front corner is cut in the illustration to present different stages in the multi-stage structure.}
\label{fig:fig3}
\end{center}
\end{figure}

The charge readout system of the LAr LEM-TPC is a direct extrapolation of the successful 
10 $\times$ 10 cm$^2$ prototype 
\cite{Badertscher:2010zg} 
to an active area as large as 
76 $\times$ 40 cm$^2$.
It is characterized by (1) a single 1-mm-thick LEM amplification stage and (2) a 2D readout anode realized on a single PCB. 
The system has a multi-stage structure consisting of two extraction grids, a LEM, a 2D readout anode and two signal collection planes as illustrated in 
Figure~\ref{fig:fig3}. 
The distances between the stages and typical configurations for the potentials and the inter-stage electric fields are summarized in Table~\ref{tab:table1}. 
In typical configurations a positive HV of $\sim$1 kV is applied to the anode while the extraction grid in liquid is operated at a negative HV of $\sim-7$ kV,
resulting in a virtual ground within the LEM plane. 

\begin{table}[hbtp]
\caption{LEM charge readout system configurations.}
\label{tab:table1}
\begin{tabularx}{\textwidth}{@{\extracolsep{\fill}}lccc} 
\hline
& Distance to the & Typical operating & Field to the stage \\
& stage above (cm) & potential (kV) & above (kV/cm) \\
\hline
\normalsize
Signal collection plane 2 & -- & 0 & -- \\
Signal collection plane 1 & 2 & $+$1 & -- \\
2D anode & 1 & $+$1 & -- \\
LEM (top electrode) & 0.2 & $+$0.5 & 2.5 \\
LEM (bottom electrode) & 0.1 & $-$3 & 35 \\
Extraction grid in GAr & 1 & $-$4 & 1 \\
Extraction grid in LAr & 1 & $-$7 & 3 \\
\hline
Cathode & 60 & $-$31 & 0.4 \\
\hline
\end{tabularx}
\end{table}

The LEM is a macroscopic hole electron multiplier built with standard PCB techniques\footnote{It was manufactured at ELTOS Circuiti Stampati Professionali,
Italy (\url{http://www.eltos.com/}).}. 
It is a 1-mm-thick FR4 plate, double-side cladded with a passivated copper layer. 
Of the order of half a million holes 500 $\mu$m in diameter are CNC (Computer Numerical Control) drilled through the plate at a pitch of 800 $\mu$m between the centers of adjacent holes. After the PCB has been manufactured and drilled, the copper layers are further chemically etched to create 40~$\mu$m wide rims
around the holes.
A HV of typically 3.5 kV is applied across the two faces creating a strong electric field of 35 kV/cm in the LEM holes leading to multiplication of electrons by avalanches. 

The 2D anode plane has two orthogonal sets of readout strips on the bottom face providing two independent readout views. 
On top of the strips for one view, those for the other view are laid with a thin (50 $\mu$m) polyimide (Kapton) insulating bed underneath. 
For each view, copper strips are formed by etching at a pitch of 600 $\mu$m covering nearly the full area of 
76 $\times$ 40 cm$^2$.
The readout pitch of 3 mm is obtained by bridging five strips at one end. 
The strips on top are narrower (120~$\mu$m) than those lying underneath (500 $\mu$m) optimizing for an equal charge collection by the two views operating at the same potential. 
A more detailed description of the 2D anode can be found in \cite{Badertscher:2010zg}. 
The strips for both of the views are tilted by 45$^{\circ}$ 
with respect to the sides of the rectangular plate,
to have symmetric conditions for both views.

We use two signal collection planes, between which 
HV decoupling capacitors (270 pF) are connected.
Each readout channel is routed to a collective 32-channel connector to a flat cable. 
A surge arrester  protecting the readout 
electronics from discharges is mounted for each channel. 
The cables bring the signals through a custom-made feedthrough on the upper flange, 
connected to the readout electronics \cite{Devis:electronics} placed outside of the vessel. 

The LEM charge readout system is assembled as a multi-layered ``sandwich'' unit with precisely defined inter-stage 
distances and inter-alignment. 
The entire unit can then be mounted to or dismounted from a field cage, which may have different 
configurations, e.g. drift lengths.

\subsection{ArDM cryogenic system}
\label{sec:2.2}

The double-phase operation of the LAr LEM-TPC requires a sophisticated cryogenic LAr apparatus which fulfills demanding conditions: 
(1) the LAr target must be maintained at a very high purity which corresponds to an oxygen-equivalent impurity concentration less than 1 ppb, 
(2) the pure argon volume  must be pumped down to a high vacuum ($\sim10^{-6}$ mbar) before filling with LAr
to reduce contamination of the argon due to outgassing, 
(3) filling through a purification filter is obligatory, 
(4) to maintain and improve the purity during the test period (i.e. months), LAr must continuously be cleaned 
with the aid of a recirculation/purification system, 
(5) thermodynamical conditions, i.e. temperatures and pressures must continuously be controlled and 
(6) the system must be protected from possible hazards due to a large amount of cryogenic LAr in case of unexpected incidents. 
Therefore, we carried out the test experiment fully exploiting the cryogenic apparatus of the ArDM Experiment, which we had
previously built and operated at CERN~\cite{Marchionni:2010fi}. 

\begin{figure}[hbtp]
\begin{center}
\includegraphics[width=0.99\columnwidth]{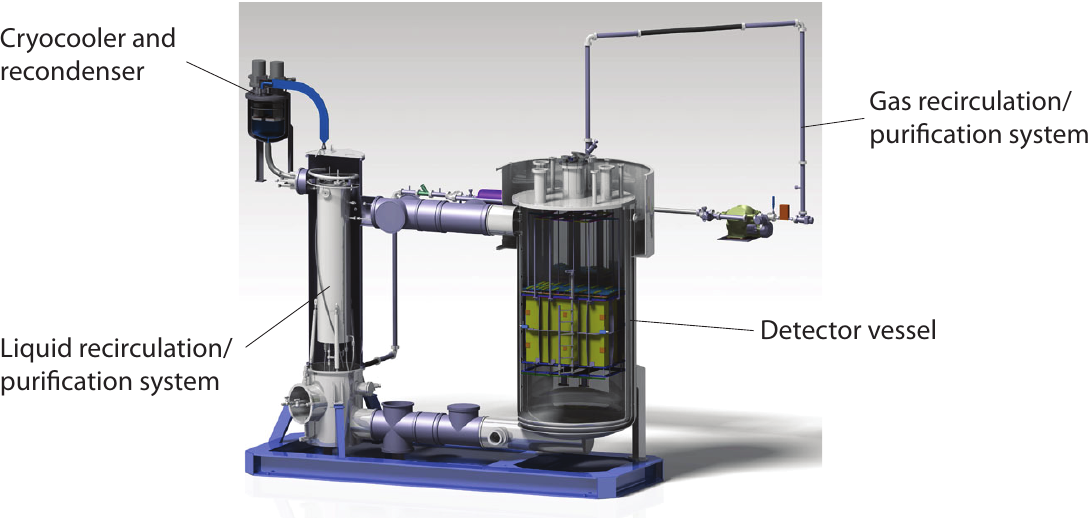}
\caption{Overview of the cryogenic system of ArDM.}
\label{fig:fig4}
\end{center}
\end{figure}

Figure \ref{fig:fig4} presents a rendered CAD picture of the apparatus.
The ArDM cryogenic system consists mainly of the detector vessel and the LAr recirculation/purification system. 
The two sections are connected with horizontal pipings at the top and the bottom. 

The detector vessel has an inner diameter of 1 m and a height of 
2 m containing approximately 2 tons of LAr when it is full. 
In this test experiment the vessel was filled up to approximately two thirds of the full volume. 
The pure LAr volume, which is constructed as a ultra-high vacuum tight closed volume, is surrounded by a separate LAr volume serving as a cooling bath. 
The temperature of LAr and consequently the pressure in the cooling bath are maintained with the aid of the cryocoolers installed in the recondenser sitting on top of the recirculation/purification section. 
The cooling bath then is surrounded entirely by an insulation vacuum. 
The pure LAr is recirculated through a custom-made purification cartridge filled with activated copper powder with the aid of a cryogenic bellows LAr pump installed below the cartridge, and is purified continuously, i.e. the residual oxygen contamination is removed through a reaction ${\rm O}_2 + 2{\rm Cu} \rightarrow {\rm 2CuO}$.  
For initial filling of the vessel with LAr another custom-made cartridge of a similar type with a molecular sieve and an active copper powder
layer was used. 
In addition a gas recirculation system incorporating a commercial SAES getter\footnote{SAES Getters S.p.A.. \, \url{http://www.saesgetters.com/}} is implemented in the system. 

The thermodynamical conditions, i.e. pressure and temperature, at different parts of the system are monitored and logged continuously by a PLC (Programable Logic Controller) system. 
Furthermore the PLC system allows manipulation of various active components such as vacuum pumps, electro valves, the bellows pump and heaters to regulate the cooling power of the cryocoolers.
Although the cryogenic system has a number of rupture disks as an ultimate safety measure 
the PLC system provides early alarms and a range of interlocks to stabilize the conditions before reaching the rupture 
pressure.
In addition, the system can be accessed via internet allowing a remote control. 

During the test experiment, the LAr temperature was maintained at 86.3~K. 
We reached a pressure stability better than $\pm 1$~mbar.
With the electron drift velocity of 1.4 mm/$\mu$s at 86.3 K and the drift electric field of 0.4 kV/cm  
it takes 430 $\mu$s for electrons to travel over the full drift length of 600 mm of the tested LAr LEM-TPC. 
We ran the bellows pump for two weeks at full speed and the oxygen-equivalent impurity 
concentration reduced from $\sim$1.5 to $\sim$0.5 ppb. 
The lifetime of drifting electrons ($\tau$) can approximately be calculated as a function of the oxygen-equivalent impurity concentration ($\rho$) \cite{Buckley:1988qx} : 
$\tau\  [\mu{\rm s}] \approx 300/\rho\ [{\rm ppb}]$. 
The impurity concentration of 0.5 ppb leads to $\tau \approx 600$ $\mu$s. 

\subsection{HV system}
\label{sec:2.3}
\subsubsection{Cryogenic Greinacher HV multiplier}
\label{sec:2.3.1}

\begin{figure}[hbtp]
\begin{center}
\includegraphics[width=\columnwidth]{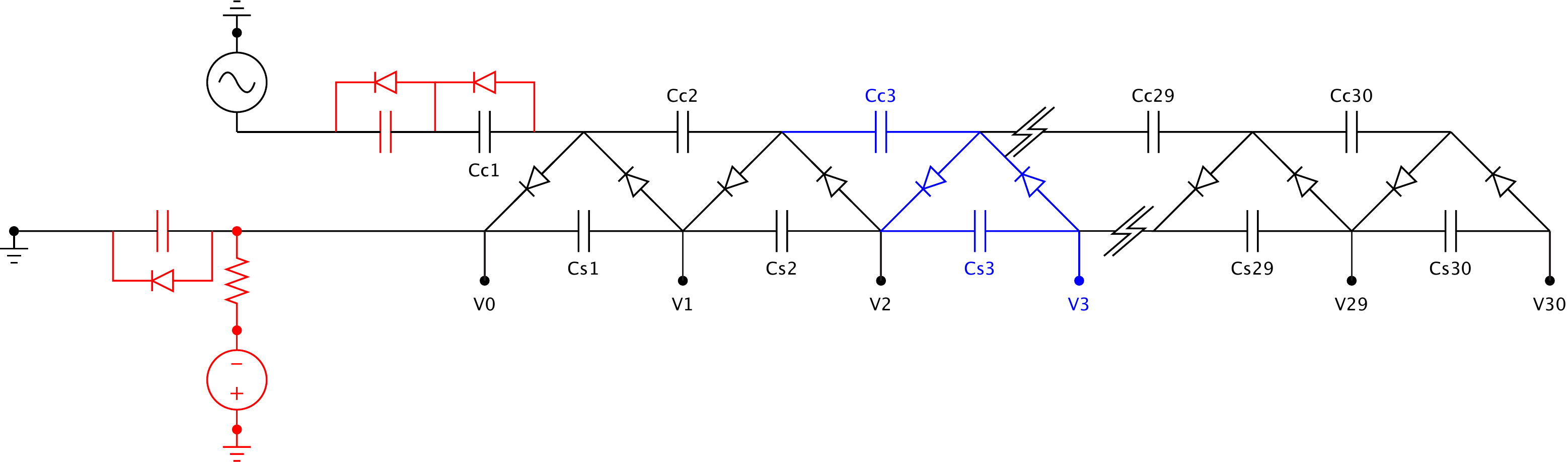}
\caption{Diagram of a 30-stage negative Greinacher circuit. 
The stage No. 3 is colored in blue to visualize the structure of a single stage. The DC voltage source, the capacitors and the diodes shown in red are used to shift the potential of the whole circuit by a certain value with respect to the ground.}
\label{fig:fig5}
\end{center}
\end{figure}

The 30-stage cryogenic Greinacher HV multiplier was designed fully exploiting experiences gained in R\&D of the ArDM HV system 
\cite{Horikawa:2010bv},
which is based on 
the same concept and has 210 stages. 
For the main circuit components, namely capacitors and diodes, ones of the same types as in ArDM were chosen because of their known reliability in LAr. 
Figure  \ref{fig:fig5} illustrates the diagram of the circuit. 
Each capacitor symbol corresponds to one single polypropylene capacitor having 82 nF and each diode symbol to a chain of three avalanche diodes connected in series. 
The direction of the diodes determines the polarity of the DC output voltage which is negative in this case.
A single stage is designed to 
hold as high as 2 kVDC, i.e. the circuit is able to output a total of $-$60 kVDC. 

The output voltage is proportional to the amplitude of sinusoidal input voltage. 
Fully charged the voltage across each stage reaches ideally the peak-to-peak value of the input voltage ($V_{\rm pp}^{\rm in}$) and hence the potential at the $i$th stage equals to  $V_i = i\cdot V_{\rm pp}^{\rm in}$. 
The DC output of the $i$th stage which is denoted as $V_i$ in the diagram is connected to the field shaper FS$i$ ($i = 0...30$). 
The potential of the whole circuit can be shifted by adding a DC voltage source 
and several appropriate components,
as shown in red in Figure~\ref{fig:fig5}. This is used to adjust the potential of 
FS0 and of the extraction grid in LAr with respect to that of the 
charge readout system. 

The Greinacher circuit is integrated directly on the outer surface of one of the side-wall PCBs 
as can be seen in Figure~\ref{fig:fig2}.
The shifting part is built on a separate PCB and is mounted 
near the Greinacher circuit.
For the last 12 stages a HV resistor having 
200 M$\Omega$ is inserted between the Greinacher stage and the field shaper to limit the current flowing out of the large capacitances in case of an unwanted discharge 
from the cathode or the field shapers at very HVs. 
For the double-phase operation mode the whole circuit is immersed in LAr. 

Table~\ref{tab:table2} summarizes the parameters of the Greinacher circuit as described above. 
The specifications of its components 
can be found in Table~\ref{tab:table3}.

\begin{table}[hbtp]
\caption{Specifications of the Greinacher circuit.}
\label{tab:table2}
\begin{tabularx}{\textwidth}{@{\extracolsep{\fill}}p{50mm}ccc} 

\hline
 \multicolumn{2}{@{}l@{}}{\rm Number of Greinacher stages}, $N$ & 30 & -- \\
 \multicolumn{2}{@{}l@{}}{\rm Capacitance per capacitor, $C$} & 82 & nF \\
 \multicolumn{2}{@{}l@{}}{\rm Number of capacitors} & 60 & --\\
 \multicolumn{2}{@{}l@{}}{\rm Combined capacitance} & 5.5 & nF \\ 
\hline
 \multicolumn{2}{@{}l@{}}{\rm Maximum rated voltage per stage} & 2000 & V\\
 \multicolumn{2}{@{}l@{}}{\rm Maximum output voltage ($V_{\mathrm{max}}$)} & 60 & kV\\
\hline
 \multicolumn{2}{@{}l@{}}{\rm Stored charge at $V_{\mathrm{max}}$} & 10 & mC \\
 \multicolumn{2}{@{}l@{}}{\rm Stored energy at $V_{\mathrm{max}}$} & 10 & J \\
\hline
\end{tabularx}
\end{table}

\begin{table}[hbtp]
\caption{Components of the Greinacher circuit.}
\label{tab:table3}
\begin{minipage}{\textwidth}
\renewcommand\footnoterule{\relax}
\begin{tabularx}{\textwidth}{@{\extracolsep{\fill}}p{60mm}cc} 
\hline
\vspace{-1.5mm} {\bf Capacitor} \\ 
\multicolumn{3}{@{}l}{Double metallized film pulse capacitor, polypropylene dielectric} \\
\hline
{\rm Manufacturer} & Evox Rifa\footnote{Evox Rifa GmbH. 
\url{http://www.evoxrifa.fi/}} \\
{\rm Type} & PHE450 \\ 
Winding construction & Two Section \\
Lead spacing & 22.5 & mm \\
{\rm Dimensions} & 11.0 $\times$ 21.5 $\times$ 26.0 & mm$^3$ \\
Rated voltage DC & 2000 & VDC \\
{\rm Capacitance} & 82 & nF \\
\hline
\vspace{-1.5mm} {\bf Diode} \\ 
\multicolumn{3}{@{}l}{High-voltage soft-recovery rectifier (avalanche diode)} \\
\hline
{\rm Manufacturer} & Philips Semiconductors\footnote{NXP Semiconductors. \, \url{http://www.nxp.com/}} \\ {\rm Type} & BY505 \\ 
Package & Rugged glass package \\
{\bf Maximum ratings} \\
Peak reverse voltage 
($-$65 ... $+$120 $^{\circ}$C)
& 2200 & V \\
Non-repetitive peak forward current & 5 & A \\
Repetitive peak forward current & 800 & mA \\
Average forward current\footnote{Averaged over any 20 ms period at 25 $^{\circ}$C.} & 85 & mA \\
{\bf Electrical characteristics} \\
{\rm Diode capacitance (0 V)} & 2 & pF \\
\hline
\end{tabularx}
\end{minipage}
\end{table}

\subsubsection{AC power supply unit}
\label{sec:2.3.2}

\begin{figure}[hbtp]
\begin{center}
\includegraphics[width=\columnwidth]{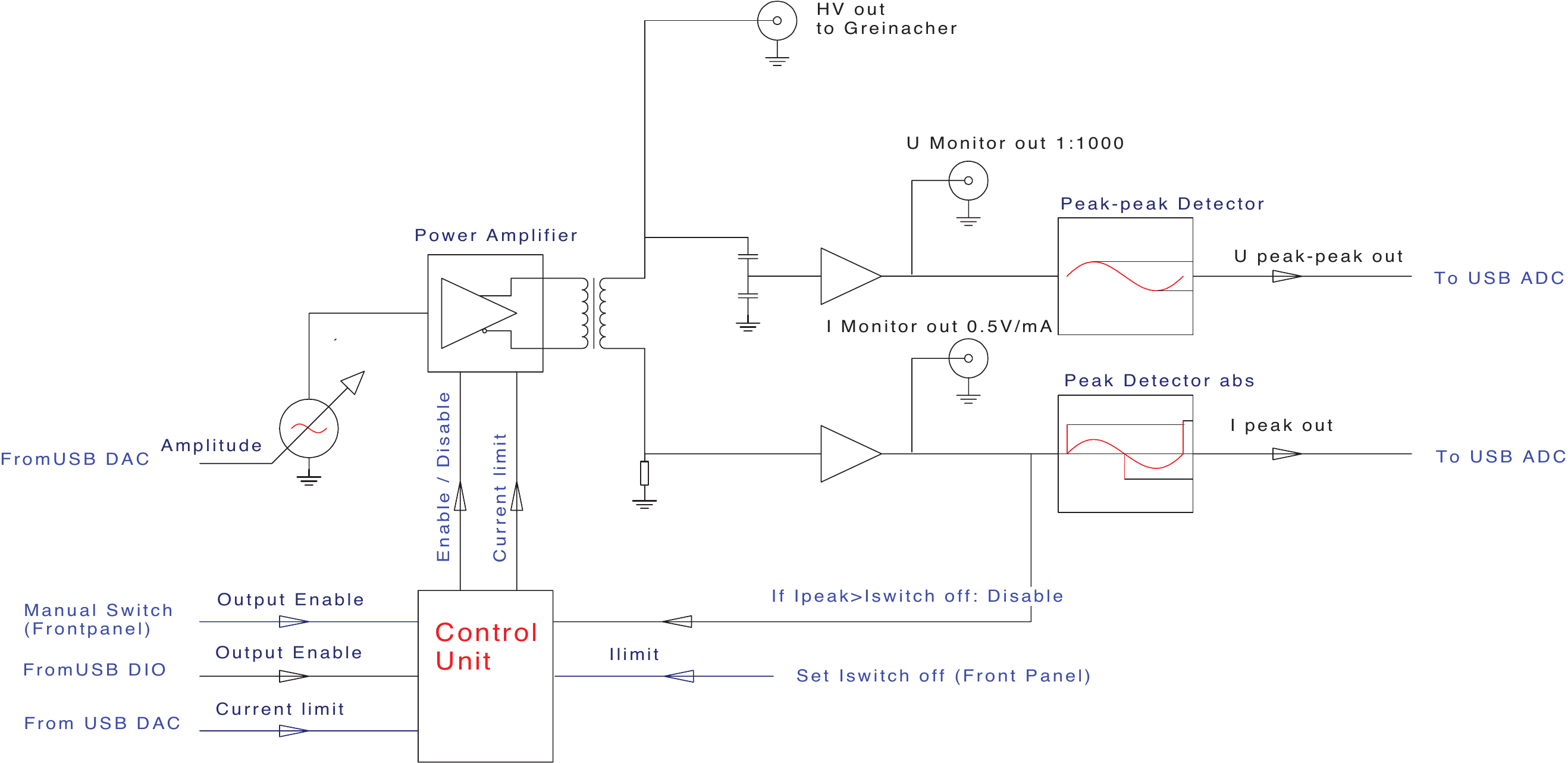}
\caption{Block diagram of the AC power supply unit for charging the Greinacher circuit.}
\label{fig:fig6}
\end{center}
\end{figure}

A dedicated power supply unit was developed to supply appropriate AC voltages for charging 
Greinacher circuits for LAr-TPCs. 
Its block diagram is shown in Figure~\ref{fig:fig6} and the relevant specifications
are summarized in Table \ref{tab:table4}.

The AC power supply unit 
is controlled and monitored through a USB interface by a commercial PC. 
The amplitude of the sinusoidal output voltage 
can be ramped up from 0 to 2.7 kV$_{\rm pp}$ (peak-to-peak) with a maximum current of 10 mA (r.m.s.). 
A fixed frequency of 50 Hz for the output signal is generated by a quartz oscillator.
The ramp-up rate of the output voltage can be set in the range from 10 to 1000 V/s. 
The output voltage and current are measured by a peak-to-peak detector and 
an absolute peak detector, respectively, and are read out via USB. 
On the front panel there is an analog output for each of the output voltage and current to allow monitoring of their waveforms on an external oscilloscope.

The internal control unit can be enabled/disabled either by the USB interface or by a manual push-button switch on the front panel. 
The peak output current is set by the 
interface through the control unit. 
If a sudden discharge of the Greinacher capacitors 
generates an over-current in the secondary coil of the transformer, the control unit turns off the system. 
This current limit 
is set by a potentiometer on the front panel. 
The power amplifier is also protected  for over-voltages. 
The system has never been damaged even though a few 
discharges occurred while testing several different Greinacher circuits. 

\begin{figure}[hbtp]
\begin{center}
\includegraphics[width=0.9\columnwidth]{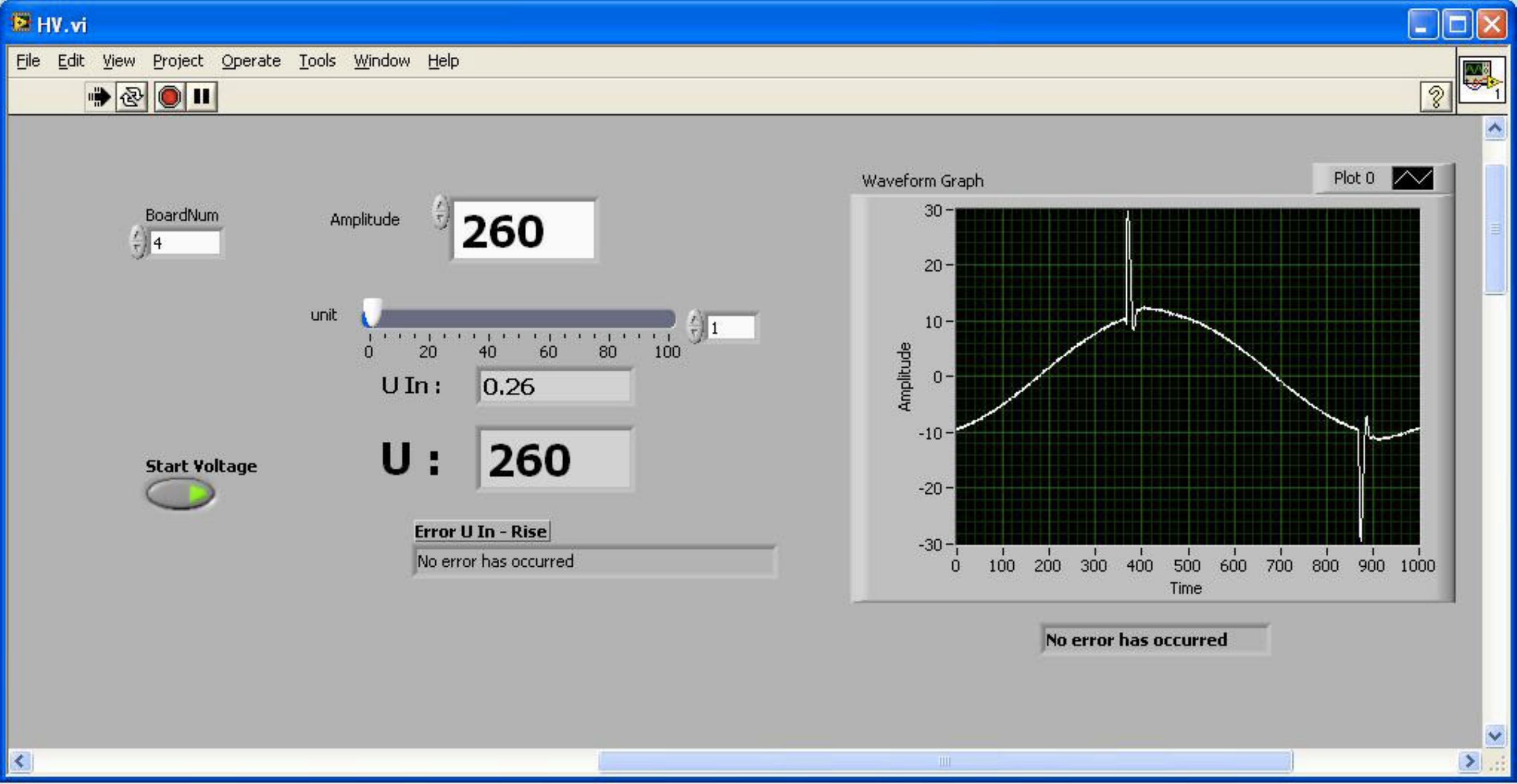}
\caption{Screenshot of the graphical interface of the Greinacher control and monitoring software.}
\label{fig:fig7}
\end{center}
\end{figure}

A 
software to control the AC power supply unit was 
built based on LabView\footnote{LabView, National Instruments Corporation. \, \url{http://www.ni.com/labview/}}. 
It has a graphical interface as shown in Figure~\ref{fig:fig7}, where one can type in the set value for the peak-to-peak output voltage and can start/stop charging by pressing a button.
The ramp-up rate 
of the output voltage can also be set using the interface.
The monitored value of the peak-to-peak voltage and a waveform of the 
output current for one cycle of the sinusoid are displayed on the window. 

\noindent
\begin{table}[htbp]
\caption{Specifications of the AC power supply unit.}
\label{tab:table4}
\begin{minipage}{\textwidth}
\renewcommand\footnoterule{\relax}
\begin{tabularx}{\textwidth}{@{\extracolsep{\fill}}p{50mm}cc} 
\hline
{\bf Output} \\
Frequency & 50 & Hz \\
Voltage & 0 ... 2700 & V$_{\rm pp}$ \\
Maximum peak current (r.m.s.) & 10 & mA \\
{\bf Transformer} \\
Manufacturer & Josef Betschart AG\footnote{Josef Betschart AG. \, \url{http://www.betschart-trafo.ch/}} \\ 
Type & TET 84a/10852 \\
Power rating & 20 & VA \\
Primary circuit (input) \\
\hspace{5mm} Maximum voltage 
(r.m.s.)
& 
30 &
V \\
\hspace{5mm} Maximum current 
(r.m.s.)
& 780 & mA \\ 
Secondary circuit (output) \\ 
\hspace{5mm} Maximum voltage 
(r.m.s.)
& 2 & kV \\
\hspace{5mm} Maximum current 
(r.m.s.)
& 10 & mA \\ 
{\bf Monitoring output} \\
Voltage monitor & 1 & V/kV \\
Current monitor & 0.5 & V/mA \\
\hline
\end{tabularx}
\end{minipage}
\end{table}

\subsubsection{Discharging system}
\label{sec:2.3.3}

\begin{figure}[hbtp]
\begin{center}
\includegraphics[height=17pc]{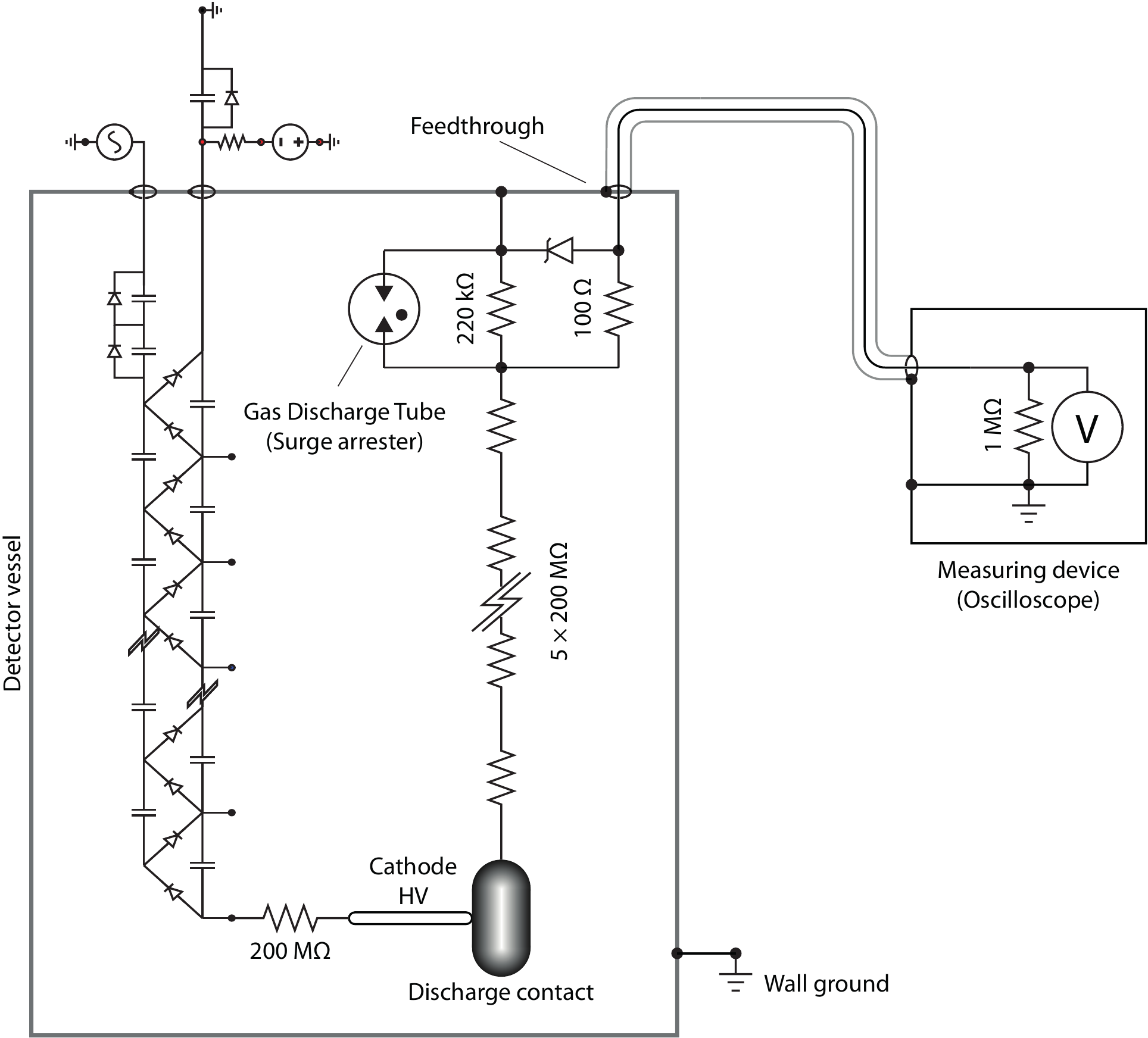}\hspace{2mm}
\includegraphics[height=21pc]{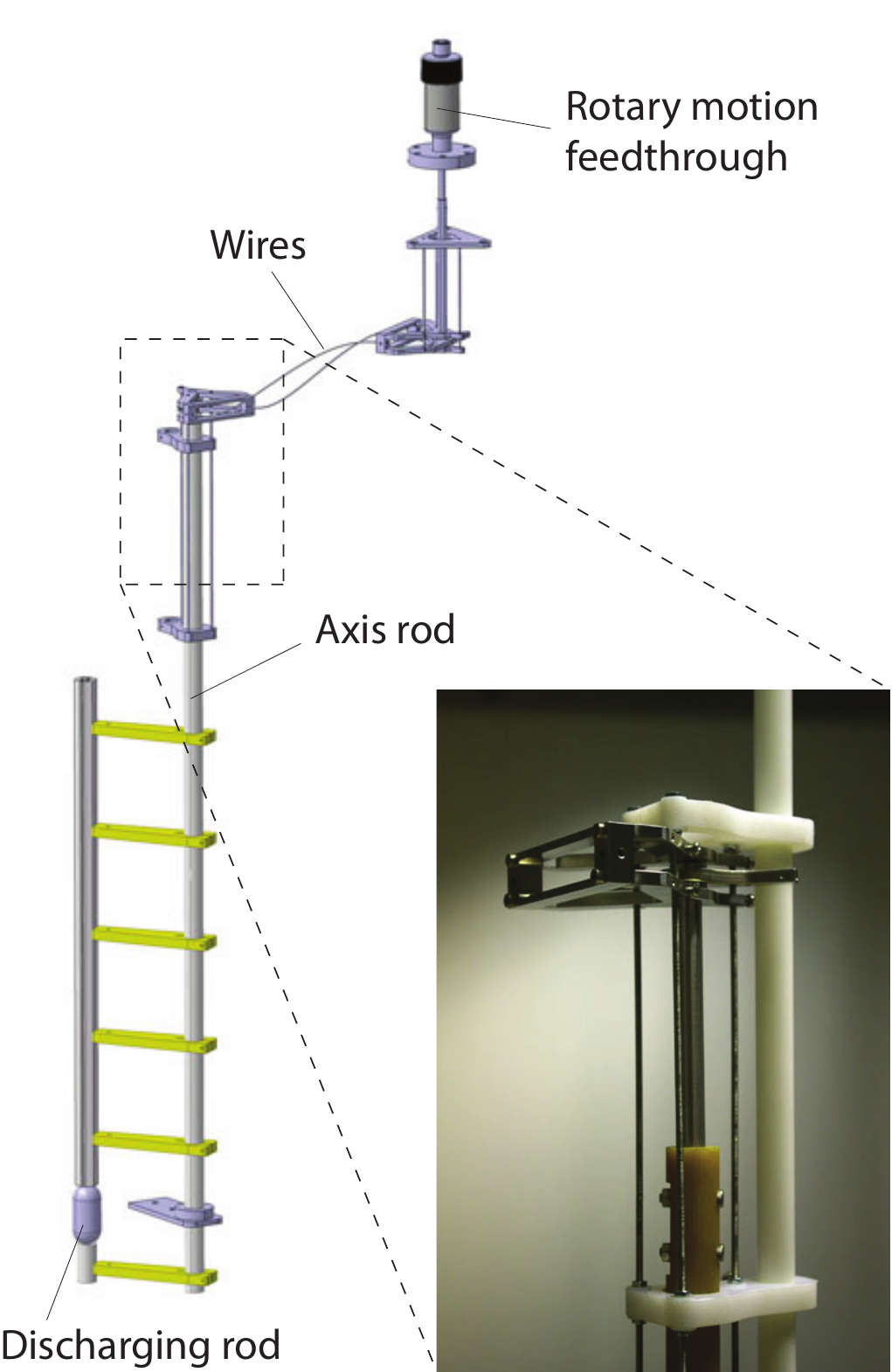}
\caption{Left: Schematic diagram of the discharging system. Discharging current can be read out by measuring the voltage across the 220 k$\Omega$ readout resistor. Right: Discharging system mechanics. The discharge contact and the resistor chain rotates about the polyethylene axis rod steered by the rotary motion feedthrough. The axis rod and the feedthrough are connected with two wires through fixed stainless steel tubes providing 
freedom to place the discharging rod independently of the feedthrough position.}
\label{fig:fig8}
\end{center}
\end{figure}

A similar system as reported in 
\cite{Horikawa:2010bv}
was built to actively and safely discharge the Greinacher circuit. 
Its circuit diagram and the mechanics are presented in Figure~\ref{fig:fig8}. 
A mechanical switch realized with a rotating discharging rod steered by a rotary motion feedthrough makes contact between the cathode and a resistor chain of total of 1 G$\Omega$ which is connected to the top flange of the dewar at the other end.
The resistance is chosen to discharge the circuit completely in the order of 10~s and to keep the peak discharging current small enough compared to the maximum rating for the forward current through the diode. 
The cathode voltage and the discharging current 
decreases with an exponential decay at a decay constant $\tau = 6.6$ s which is determined by the chain resistance $R = 1.2$ G$\Omega$ including the resistor inserted between the Greinacher circuit and the cathode and the combined capacitance $C = 5.5$~nF  of the Greinacher circuit. 
The peak current for discharging from 60 kV is 50 $\mu$A. 

The newly developed mechanics have a cable-pull system like a bicycle brake, i.e. the axis rod and the feedthrough shaft are connected with two stainless steel wires 
through two fixed stainless steel tubes.
Thanks to this new technique the wires can be routed flexibly providing a large freedom in choosing the location of the discharging rod independent of the position of the feedthrough. 

Such a discharging system allows not only discharging the circuit but even more importantly a direct measurement of the cathode voltage by measuring the peak value of the discharging current. 
The discharging current can be measured by recording the voltage across the 220 k$\Omega$ reading resistor inserted between the discharging resistor chain and the ground. 
The measured peak voltage is equal to 
0.15 per mil of the cathode voltage, e.g. 1.5 V is measured for 10 kV. 
A surge arrester and a Zener diode are connected in parallel to the reading resistor to protect the measuring device, which is usually an oscilloscope. 

\section{Characterization of the Greinacher HV multiplier}
\label{sec:3}

\subsection{Theoretical model}
\label{sec:3.1}

The total output voltage expected from a Greinacher chain is ideally equal to the value of the input peak-to-peak AC voltage multiplied by the number of stages. 
Under real conditions however, the output voltage can be smaller. Losses are mainly due to effective capacitances of the diodes 
and several stray capacitances which depend on the actual geometrical setup.
Altogether the ``shunt'' capacitance  is  the capacitance between the AC side (the top row in the diagram in Figure~\ref{fig:fig5}) 
and the DC side (the bottom row) of the Greinacher circuit. 
In a transmission line model, which can be found in \cite{Weiner, Everhart} and also in \cite{Horikawa:2010bv} 
the joint effects of charge dissipation in an external load resistance $R_L$ and charge in the shunt capacitances 
can be calculated to yield
the total output voltage, given as 
\begin{equation}
V^{\rm out} = V_{\rm pp}^{\rm in} \cdot N \cdot \frac{{\rm tanh}(2\gamma N)}{2 \gamma N},
\ \ \mathrm{where} \ \gamma \approx \sqrt{\frac{N}{2\nu R_LC_{\rm s}}+\frac{C_{\rm p}}{C_{\rm s}}}. 
\label{eq:slope}
\end{equation}
$N$ is the number of stages, $\nu$ the frequency of the input voltage generator,
$R_L$ the resistive load (assumed to be equally distributed on each stage),
$C_{\rm p}$ the shunt capacitance (parallel to each diode) and 
$C_{\rm s} (\gg C_{\rm p})$ the capacitance of a single capacitor element. 
The slope of the Greinacher output equals to $V^{\rm out}/V_{\rm pp}^{\rm in} =  {\rm tanh}(2\gamma N)/2\gamma $.
Neglecting the resistive load contribution,  
the  shunt capacitance can be estimated directly from it. 

In the actual ArDM setup, where the Greinacher is connected to the field shaping rings and
the cathode, we can expect surface leakage currents which can affect the response
of the Greinacher circuit, acting as a $R_L$ load. 
We first performed a set of measurements in air and in liquid
argon with the circuit disconnected from the ArDM setup, as described in the next Section.

\subsection{Charging up of the circuit at room and LAr temperatures}
\label{sec:3.1b}

Before assembling the whole TPC we tested the Greinacher circuit alone -- not connected to the field cage --
in air at room temperature and in a polystyrene container open to the atmosphere and filled with LAr
for characterizing the circuit. 
For measuring the output HV we used a custom-made field mill \cite{Kaufmann2008,fieldmill}, 
which is a capacitive device allowing measuring the output voltages of the Greinacher circuit without 
drawing any current. 
Such a feature is very important since a resistive load resulting in a small current
connected to the Greinacher circuit changes its behavior. 

\begin{figure}[hbtp]
\begin{center}
\begin{minipage}[t]{0.47\columnwidth}
\begin{center}
\includegraphics[height=17pc]{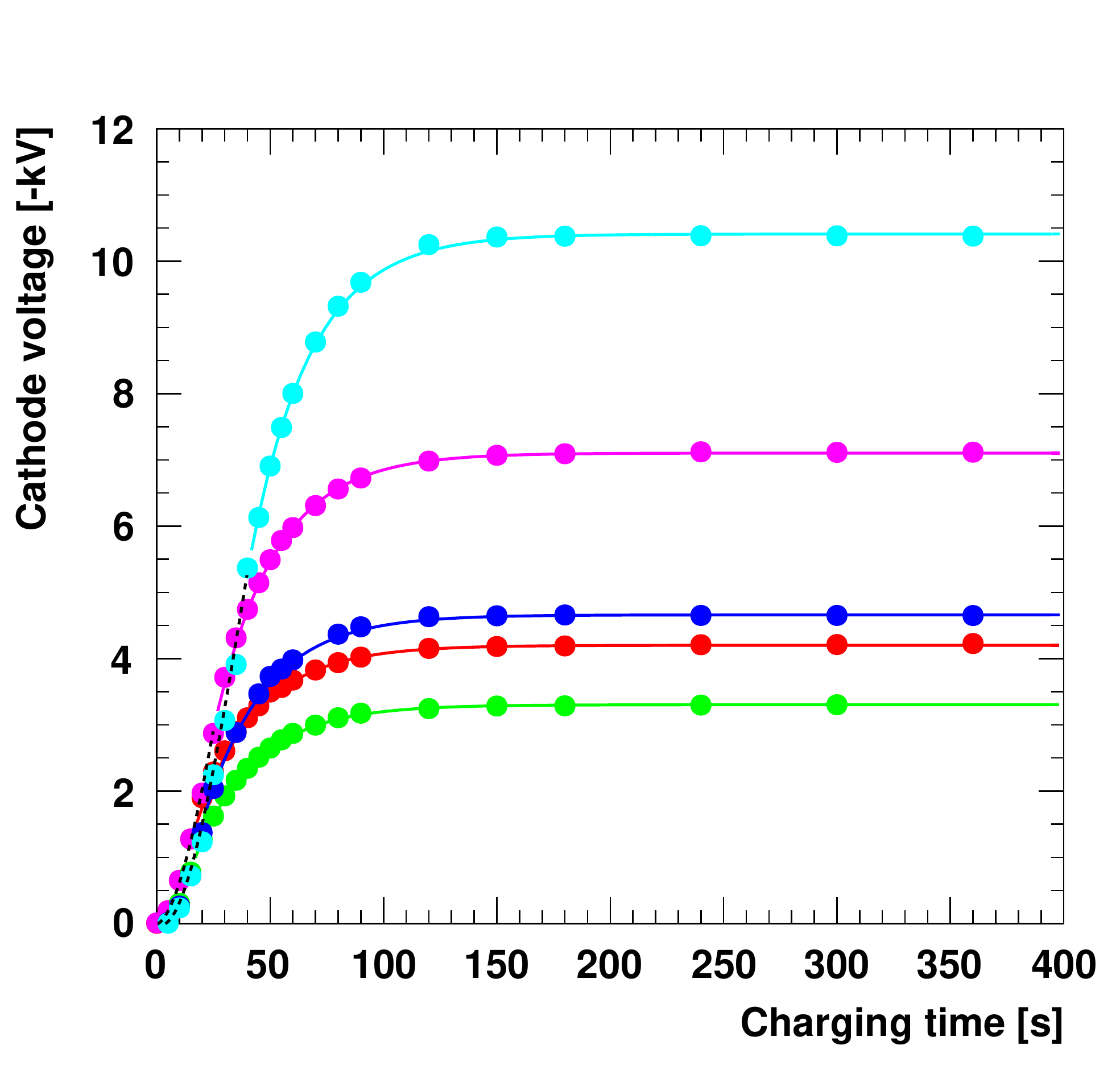}
\end{center}
\end{minipage}\hspace{2pc}%
\begin{minipage}[t]{0.47\columnwidth}
\begin{center}
\includegraphics[height=17pc]{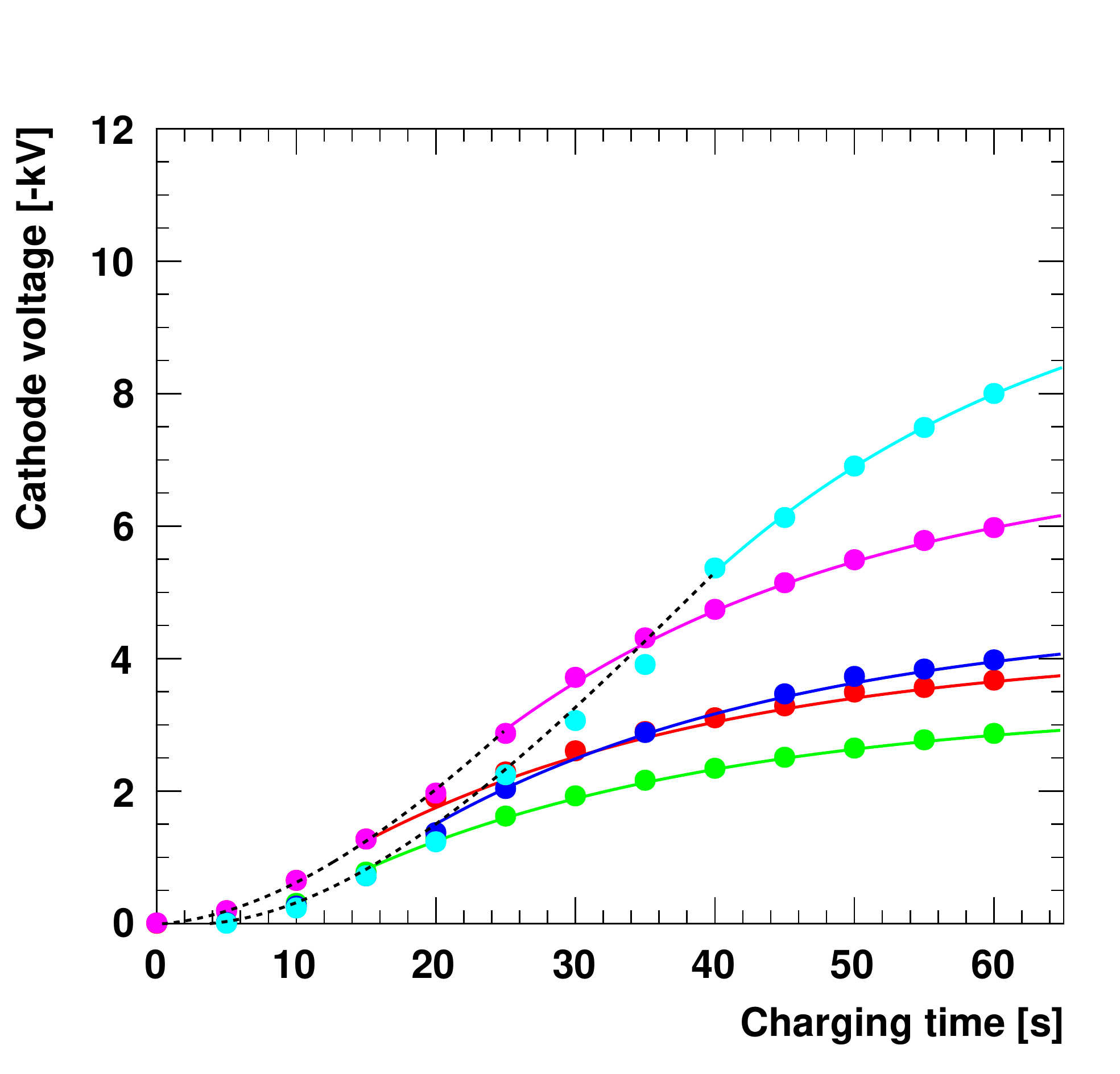}
\end{center}
\end{minipage}%
\caption{\label{fig:fig9}
Cathode voltage recorded as a function of time measured in air at room temperature or in LAr while the Greinacher circuit was charging. Different curves are for different input AC voltages, i.e. 150 V$_{\rm pp}$ (red) and 250 V$_{\rm pp}$ (magenta) in air and 150 V$_{\rm pp}$ (green), 200 V$_{\rm pp}$ (blue) and 400 V$_{\rm pp}$ (cyan) in LAr. The right plot is a zoom for the first 65 s. The dotted curves show the trends while the amplitude of the input AC voltage was ramping up (10 V/s). The solid curves show the exponential function obtained in Figure~\protect\ref{fig:fig10}.} 
\end{center}
\end{figure}

Figure \ref{fig:fig9} shows thus measured output voltage at the cathode (i.e. the Greinacher stage No. 30) while the circuit was charging up. 
Different curves are for different input AC voltages, i.e. 150~V$_{\rm pp}$ (red) and 250 V$_{\rm pp}$ (magenta) in air and 150 V$_{\rm pp}$ (green), 200 V$_{\rm pp}$ (blue) and 400 V$_{\rm pp}$ (cyan) in LAr. 
The horizontal axis shows the time after the AC voltage source was turned on. 
The input AC voltage 
was 
ramped up at a fixed rate of 10 V/s throughout the operation of the system. 
The trend of the increase of the voltage at the first part of each curve thus is determined by the ramp-up rate and is independent of the input voltage, as can be seen clearly in the zoom for the first 65 s shown 
in the right plot in Figure~\ref{fig:fig9}. 
This trend however is different between in air and in LAr due to the threshold effect of the diodes in cold (see also Section~\ref{sec:3.2}). 
For each of the two temperatures it was parametrized with a polynomial function and is shown with a dotted curve. 

\begin{figure}[hbtp]
\begin{center}
\includegraphics[height=17pc]{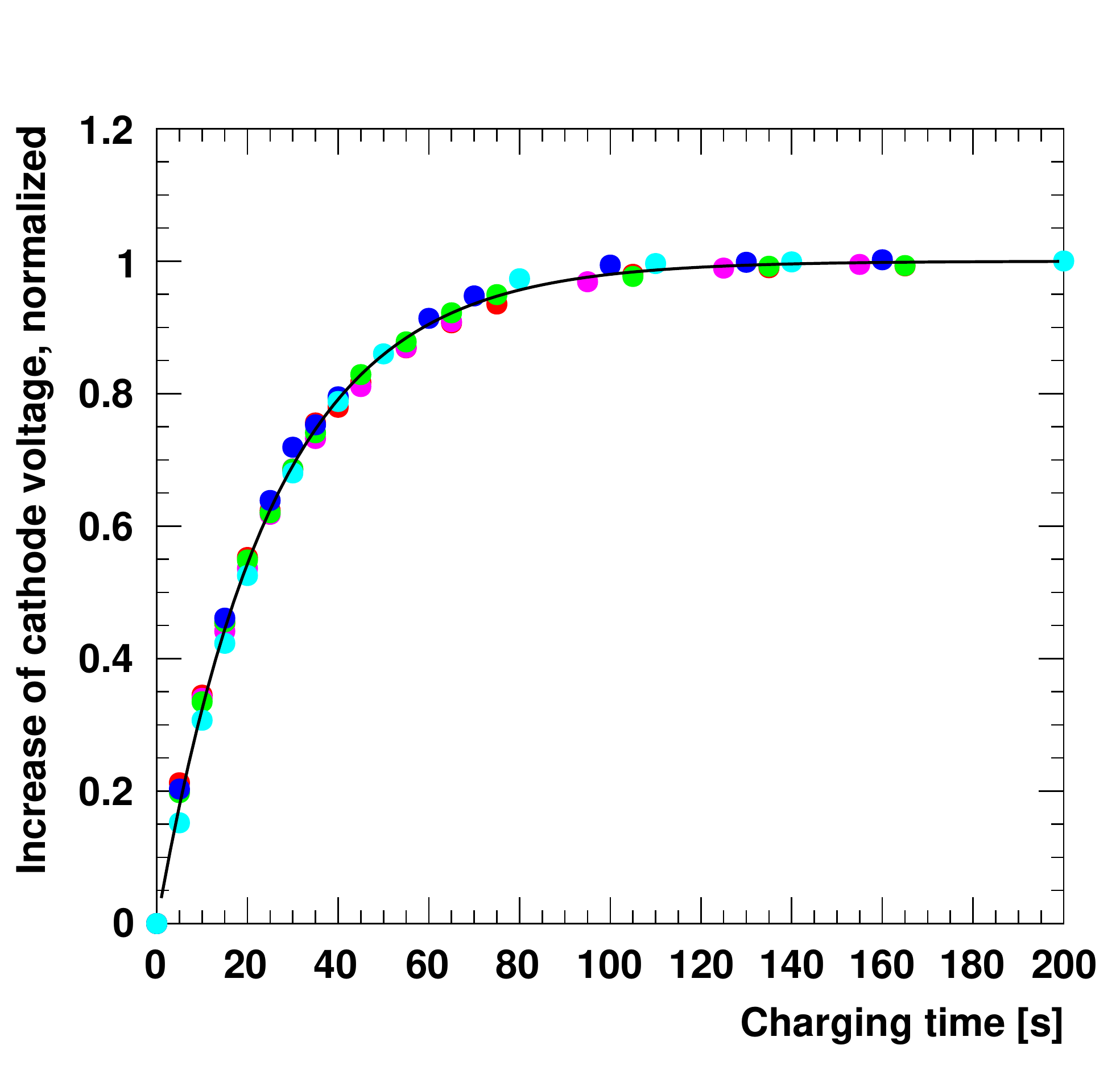}
\end{center}
\caption{\label{fig:fig10}
Increase of the cathode voltage after the amplitude of the input AC voltage reached the set value. Different colors correspond to those in Figure~\protect\ref{fig:fig9}. All the points agreed with a single exponential function shown with the solid line regardless of the input voltage or the temperature. The time constant was found to be $\sim$26 s.} 
\end{figure}

To see a general trend of charging of the circuit discarding the effect of the ramp-up rate of the AC power supply, 
the increase of the output voltage after the set value of the input voltage was reached is plotted in Figure~\ref{fig:fig10}. 
The increase of the voltage is normalized to the difference between the voltage measured when the set value was reached and the finally reached voltage. 
Thus all the curves shown in Figure~\ref{fig:fig10} lie on a single curve and fit very well with an exponential function, implying that the charge-up time was approximately independent of the input/output voltages and also of the temperature. 
Fitting all the points with an exponential function $V(t)/V_{\rm fin} = 1 - e^{-\frac{t}{\tau}}$ the time constant $\tau$ for the charge up of  the circuit was found to be 26 s, i.e. the circuit was charged up to 99\% of the final value in 120 s after the input AC voltage reached the set value. 

\subsection{Linearity of the DC output voltage as a function of the AC input voltage}
\label{sec:3.2}

\begin{figure}[hbtp]
\begin{center}
\includegraphics[height=17pc]{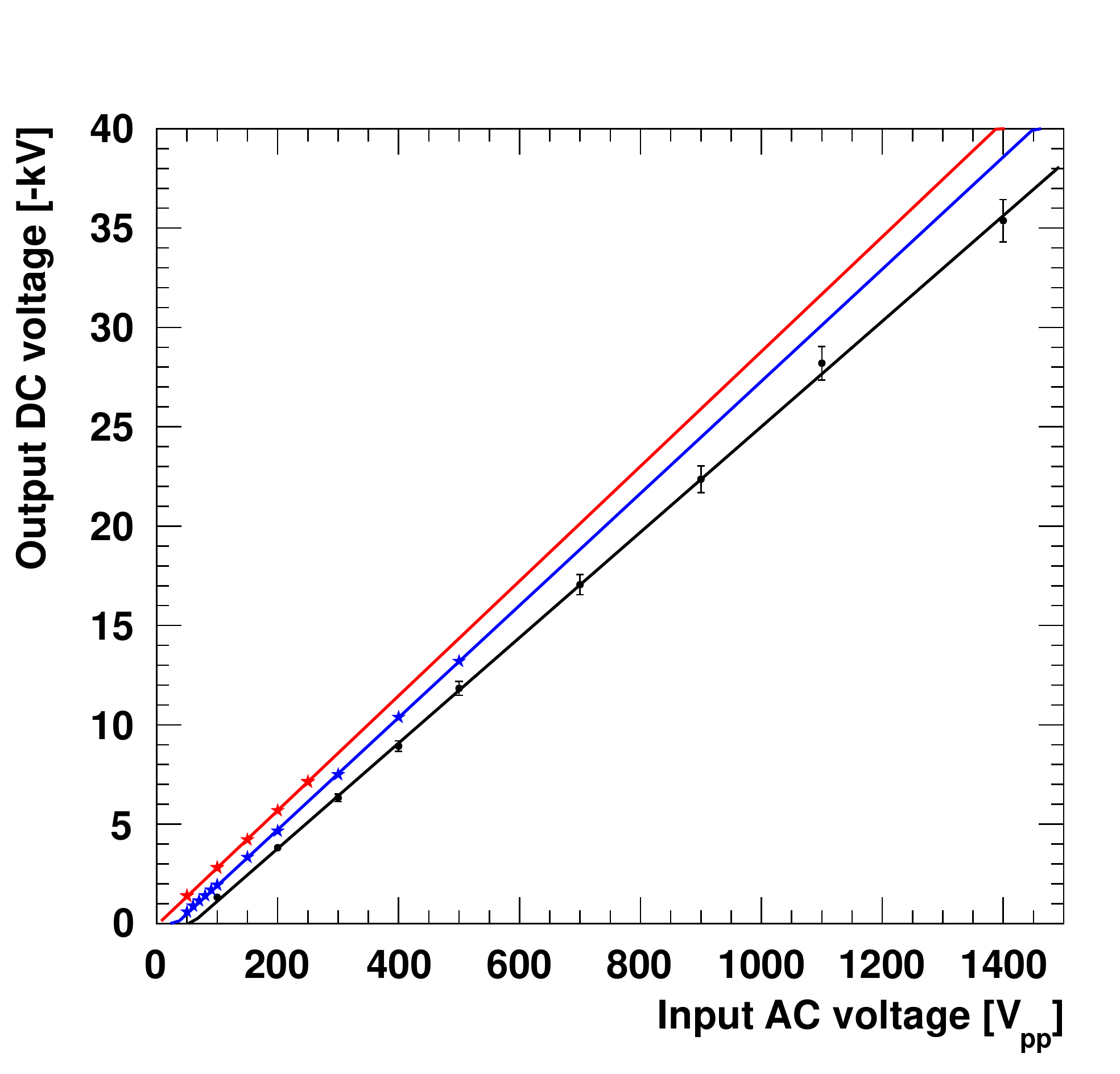}
\end{center}
\caption{\label{fig:fig11}
Output DC voltage as a function of the peak-to-peak input AC voltage measured in air (red) and immersed in LAr (blue).
Same measured in the ArDM setup using the discharging rod (black). See Section~\protect\ref{sec:4.3}. 
} 
\end{figure}

Figure \ref{fig:fig11} shows the output DC voltage measured at the last stage of amplification
as a function of the peak-to-peak input AC voltage in air at room temperature (red) and in LAr (blue). 
The points measured in air 
agreed well with a linear dependence between the output and input voltages over the whole range of the tested output voltages up to 7.5 kV. 
The curve for LAr was shifted towards higher 
input voltage by 33~V$_{\rm pp}$ showing the threshold effect of the diodes in cold, i.e. the diode drew forward current only when the voltage across the three series diodes exceeded the threshold value.  
However a linear dependence was verified also in LAr up to 13.5 kV output although at lower voltages the points showed a small deviation, 
and therefore the points below 100 V$_{\rm pp}$ were excluded from the fit. 
The slopes obtained from a linear 
fit 
were 28.9~V/V in air and  28.2~V/V in LAr, respectively. 
In the ideal case the slope equals to the number of the Greinacher stages, i.e. 30. 
The shunt capacitances  obtained from the slopes (Eq. \ref{eq:slope}) were 2.7 and 4.4~pF for in air and in LAr, respectively. 
For an identical geometrical configuration, a larger value in LAr is expected because of its relative dielectric constant of 1.5. 
The remaining difference is due to a dissimilar capacitance between the Greinacher circuit and the surrounding ground
in the two setups (air and LAr configuration).

\subsection{Linearity of the voltage distribution across the multiplying stages}
\label{sec:3.3}

\begin{figure}[hbtp]
\begin{center}
\begin{minipage}[t]{0.47\columnwidth}
\begin{center}
\includegraphics[height=17pc]{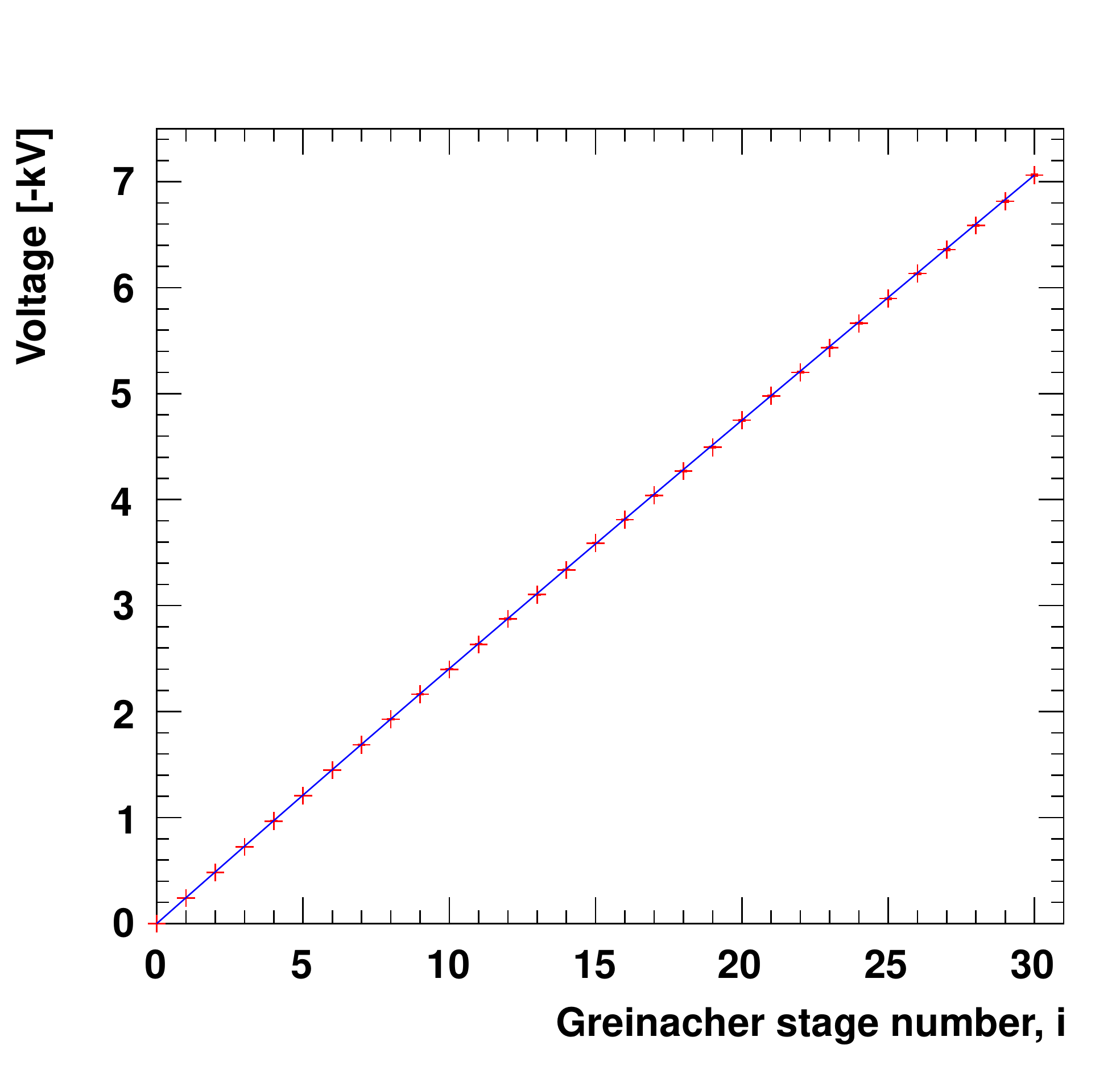}
\end{center}
\end{minipage}\hspace{2pc}%
\begin{minipage}[t]{0.47\columnwidth}
\begin{center}
\includegraphics[height=17pc]{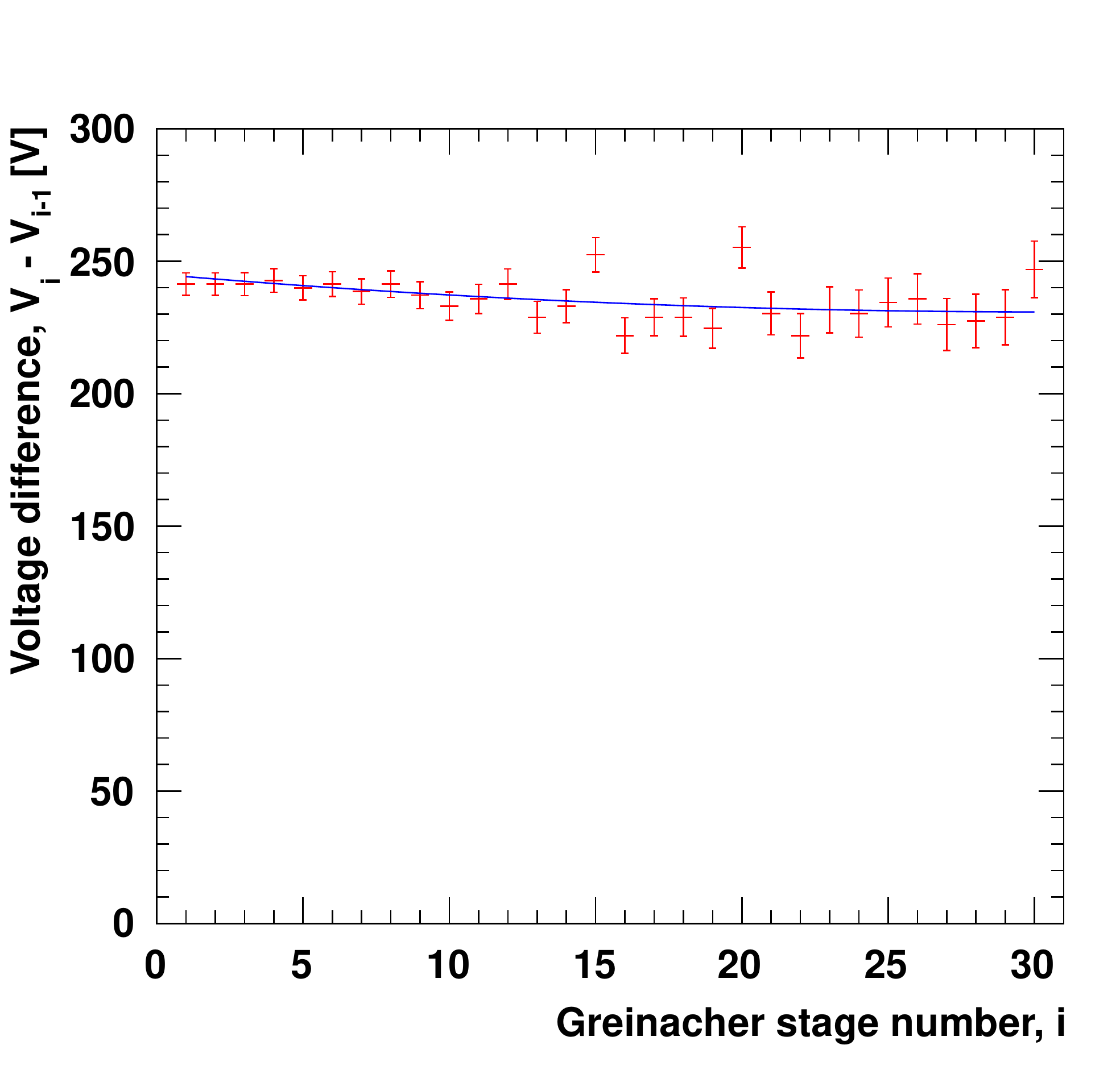}
\end{center}
\end{minipage}%
\caption{\label{fig:fig12}
Left: Measured voltage at each Greinacher stage charged at 250 V$_{\rm pp}$ in air at room temperature; Right: 
Measured voltage differences across each stage
and calculated curve (blue) using the transmission line model with a shunt capacitance. See text.} 
\end{center}
\end{figure}

The linearity of the voltage distribution over the field shapers is the first indicator of the uniformity of the drift field. 
Potentials at different Greinacher stages charged at 250~V$_{\mathrm {pp}}$ were measured in air at room temperature. 
A linear behavior of the Greinacher stage voltages was obtained, as shown in the left plot in Figure~\ref{fig:fig12}.
On the right plot, red points show the measured voltage differences 
between neighboring Greinacher stages. 
The error shown in the plot corresponds to 0.1\% of each measured voltage of field shapers. 
The blue curve is calculated using a transmission line model with a shunt capacitance of 2.7 pF as described in the previous section. 
The expected reduction of the voltage across the last Greinacher stage 
with respect to 
that of the first stage 
is $\sim$6\%.
The measured values were in good agreement with expectation although a few points deviated by $\sim$10\%. 
The expected reduction can be as large as $\sim$10\% in LAr from the calculation based on the shunt capacitance of 4.4 pF (Section \ref{sec:3.2}).

\subsection{Characterization of the HV system with the discharging method}
\label{sec:4.3}

The cathode voltage, measured by discharging the Greinacher 
circuit using the discharging system, as described in Section~\ref{sec:2.3.3}, can
be used effectively to characterize the HV system.
A series of dedicated measurements were performed with the fully assembled
setup in ArDM.

\begin{figure}[hbtp]
\begin{center}
\includegraphics[height=17pc]{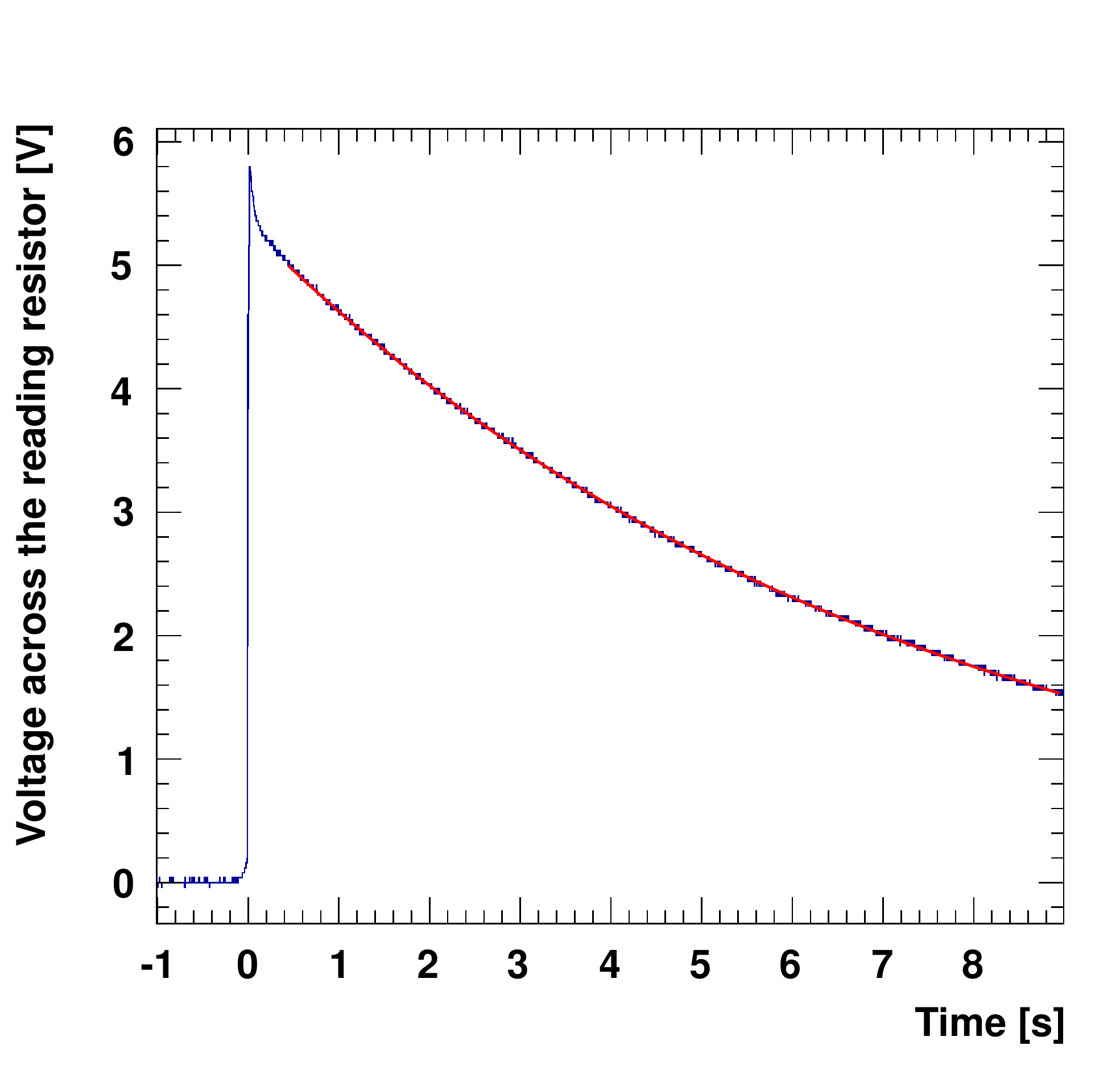}
\caption{Discharging curve at the input AC voltage 1400 V$_{\rm pp}$ fitted
with an exponential decrease of the voltage.
The intercept of the exponential $V_0 = 5.31\pm0.13$~V at $t=0$ corresponds to the cathode voltage of $35.3\pm0.9$~kV.}
\label{fig:fig17}
\end{center}
\end{figure}

A typical example of a measured  curve is shown in Figure~\ref{fig:fig17}. 
The sharp peak at $t = 0$~s 
shows the  fast discharging of the cathode grid. 
It is followed by a slower exponential decrease of the voltage due to discharging of the Greinacher circuit through the current limiting resistor (200 M$\Omega$) 
in addition to the discharging resistor chain (1 G$\Omega$), as described in Section~\ref{sec:2.3.3}. 
The discharging curve measured across the reading resistor 
was 
fitted with an exponential function $V(t) = V_0 \cdot e^{-\frac{t}{\tau}}$ for $t > 0.4$~s. 
The intercept $V_0$ is proportional to the cathode voltage before discharging. 
The shown curve was obtained for the circuit charged at 1400~V$_{\rm pp}$ input AC voltage and $V_0 =5.31\pm 0.13$~V corresponds to the cathode voltage 
of $35.3\pm0.9$~kV. 
The measured decay constant $\tau = 7.2\pm0.3$~s is consistent with the
naive calculation 
of $\tau = RC = 6.6$ s (see Section~\ref{sec:2.3.3}), taking into account
possible temperature variations of the capacitors and
additional stray-capacitance of the ArDM setup.

To measure the output voltage as a function of peak-to-peak input AC voltage 
the Greinacher circuit was charged up 
at different input voltages and then discharged recording the discharging curve. 
The measured points are plotted in black in Figure~\ref{fig:fig11}. 
Voltage measurements using the discharging current clearly had a larger error compared to those using the field mill. 
The errors were dominated by those on the resistances and on the determination of the intercept by the fit and were found to be $\sim$3\% 
of the measured voltages. 
Within the measurement precision a linearity of the output voltage to the input voltage was verified up to the input AC voltage of 1400 V$_{\rm pp}$ which resulted in the output voltage of $35.6 \pm 0.7$ kV (from the linear fit for the input voltages $\ge$200 V$_{\rm pp}$) 
and the drift electric field of $\approx$0.6 kV/cm.
The slope of the linear fit was found to be 
26.5~V/V, 
somewhat smaller than that obtained from the measurements with the Greinacher circuit alone in LAr (blue). 
The observed reduction of the slope by $\sim 6\%$ cannot be attributed to a resistive load, which
is constrained by the observed discharging time (see Section~\ref{sec:4.1}) to be $R_L>2.5~\mathrm{T}\Omega$. It can be described by two effects:
(1) the presence of the DC-voltage shifting circuit which decreases
the slope by $\sim$3\%; (2) a capacitive contribution of the dewar inner surface
and the field shapers to the shunt capacitance. 
It should be noted that the measured output voltages will be confirmed by the drift velocity measurements,
 as shown in Section~\ref{sec:4}.

We also investigated the charging curves at the input AC voltage of 500 V$_{\rm pp}$  and 900~V$_{\rm pp}$. 
Each point was measured in such a way that the input AC power supply was turned off at a certain time, while the Greinacher 
circuit was still charging, and then the circuit was discharged and the cathode voltage was measured from the discharging current. 
The behavior was found to be  similar to that shown in Figure~\ref{fig:fig9}, 
with the first part determined by the ramp-up rate of the input AC voltage followed by the second part consistent with an exponential function. 
The time constant of the exponential part was also  consistent with that obtained from Figure~\ref{fig:fig10} and showed
that the circuit was fully charged $\sim$2 minutes after the set value was reached at the input AC power supply.

\subsection{Characterization of the HV DC-shifting circuit}
\label{sec:4.4}

\begin{figure}[hbtp]
\begin{center}
\includegraphics[height=17pc]{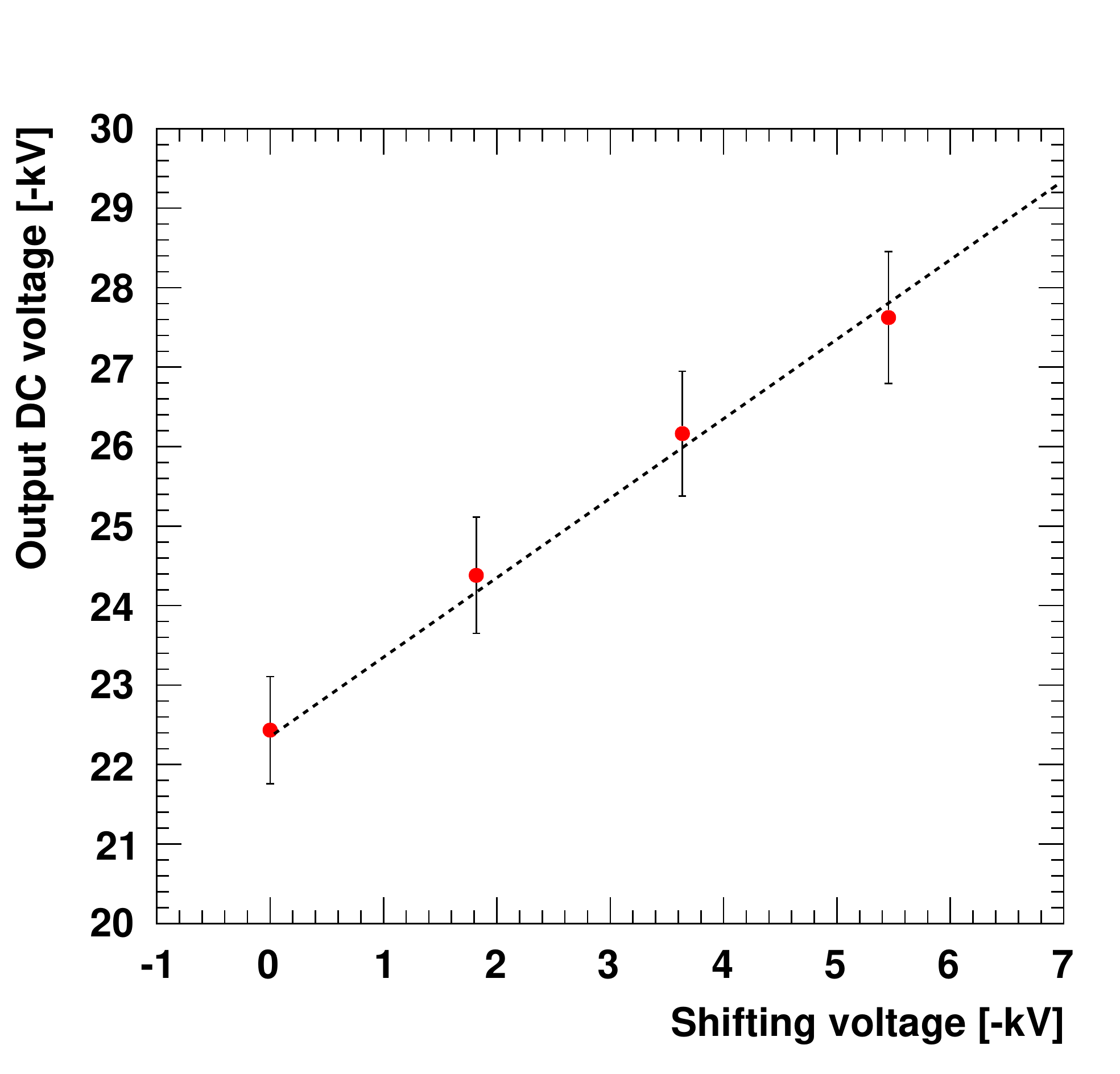}
\caption{Cathode voltage measured using the discharging rod as a function of the DC offset voltage. The error bars correspond to 
$\pm$3\% 
of the measured output voltage. The dotted curve shows the calculated value.}
\label{fig:fig19}
\end{center}
\end{figure}

To check the functionality of the DC-voltage shifting circuit 
the cathode voltage was measured with the discharging method,
while the shifting DC HV was applied at different offset values as plotted in Figure~\ref{fig:fig19}. 
The measurements agreed with the expected values (dotted curve)  verifying 
a linear increase of the cathode voltage depending on the offset voltage within the measurement precisions. 

\section{Operation of the LAr LEM-TPC in the double-phase mode}
\label{sec:4}
\subsection{Drift velocity based on scintillation light measurements}
\label{sec:4.1}

The LAr LEM-TPC was operated 
in the double-phase mode in pure argon for a continuous period of about a month,  
recording cosmic track events (a large fraction were vertically traversing muons) with the PMTs and 
the LEM charge readout system. 
The drift electric field was created using the built-in Greinacher HV system and was maintained essentially throughout the experiment. 
The Greinacher circuit was re-charged periodically. 
The recording of cosmic events could be done when the charging AC power supply was on or off. 
The natural decay
rate of the HV in the system
was measured several times during the experiment and was found to be consistent with an exponential decay with a time constant 
of the order of 5--10 days. 
Since the natural decay rate  is very slow, 
changes of the drift field were negligible for a few hours of measurements even when the AC power was off. 

\begin{figure}[hbtp]
\begin{center}
\begin{minipage}[t]{0.47\columnwidth}
\begin{center}
\includegraphics[height=17pc]{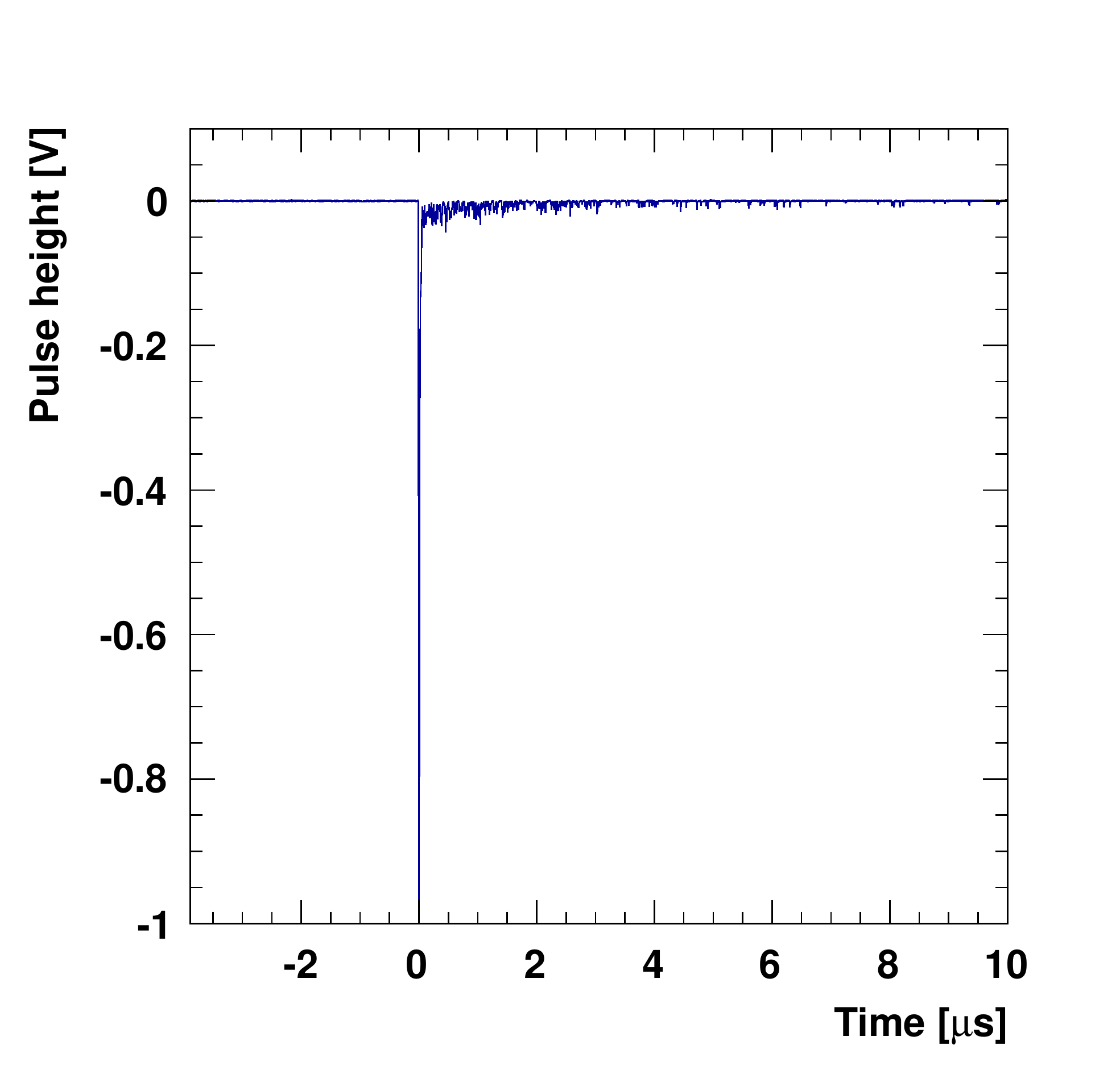}
\end{center}
\end{minipage}\hspace{2pc}%
\begin{minipage}[t]{0.47\columnwidth}
\begin{center}
\includegraphics[height=17pc]{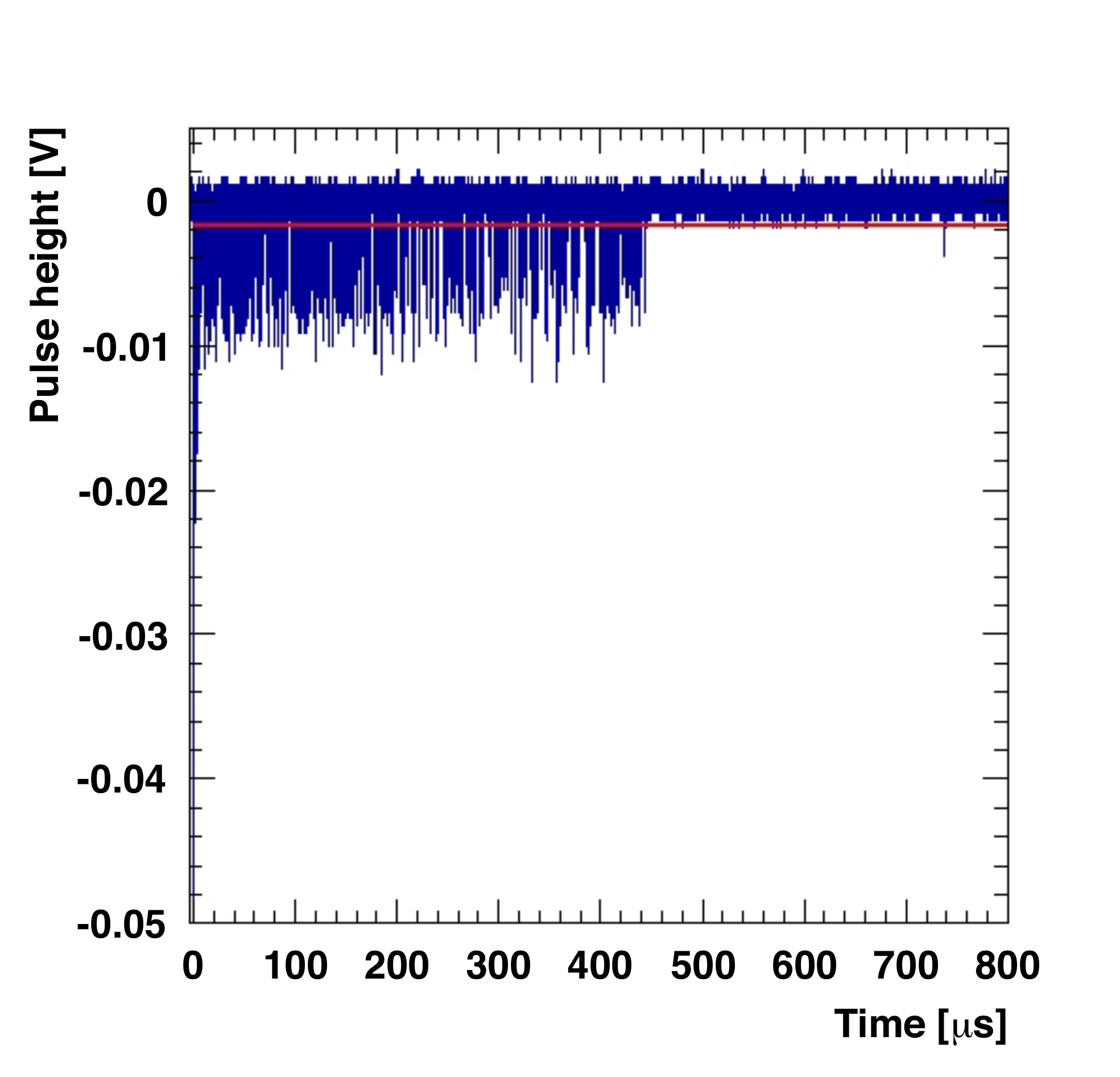}
\end{center}
\end{minipage}%
\caption{\label{fig:additional_figure}
A waveform of the PMT signal for the primary (S1, left) and secondary (S2, right) scintillation light of the same cosmic event recorded under the drift electric field of 355 V/cm. The red line in the right figure shows a typical threshold value (corresponding to $\sim$0.25 p.e.) used to discriminate PMT pulses from the noise.
The maximum drift time for this event is about 440 $\mu$s.} 
\end{center} 
\end{figure}

With the drift of electrons 
and their extraction  across the liquid-vapor interface,
signals of ionization in LAr can be detected using the PMTs through the S2 secondary scintillation light. 
It should be noted that the LAr LEM-TPC prototype was not fully optimized for  light readout.
%
The solid angle coverage by the photocathode of the PMT for the S2 light was only $\sim$0.05\%, 
 the extraction region being
located at the distance of $\sim$70 cm from the PMT, but was sufficient to detect clear primary (S1) and secondary (S2) signals,
and in particular, their time distribution. 
Typical recorded waveforms for the S1 and S2 are shown in Figure \ref{fig:additional_figure}.
The two histograms correspond to the same cosmic event taken with a drift  field of 355 V/cm. 
The S1 signal is used to determine the event time ($T_0 = 0$ $\mu$s). 
The S2 signal consists of single photoelectron pulses distributed over several hundred microseconds, 
corresponding to the different drift times of the recorded hits.
The red line in the right figure shows a typical threshold value (corresponding to $\sim$0.25 p.e.) 
used to discriminate PMT pulses from the noise.

\begin{figure}[hbtp]
\begin{center}
\includegraphics[width=0.7\columnwidth]{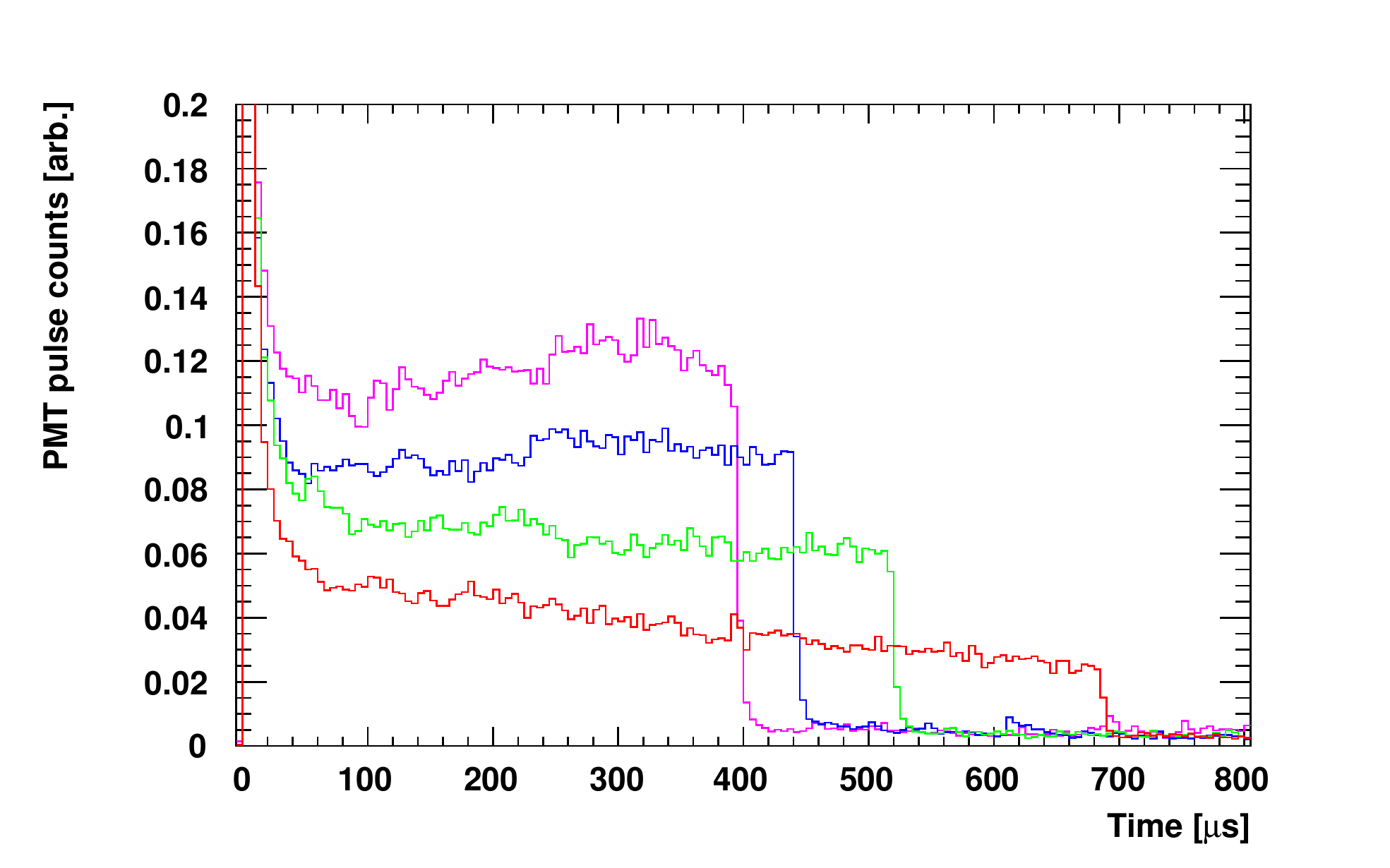}
\caption{Time distribution of PMT signals with a presence of the extraction of electrons across the liquid-vapor interface for four different electric drift fields, i.e. 
443 V/cm (magenta), 355 V/cm (blue), 266 V/cm (green) and 178 V/cm (red). 
The sharp peak at $t = 0$ $\mu$s is the primary scintillation (S1) followed by continuous S2 signal. The S2 signal shows a clear edge at the maximum drift time ($t_{\max}$) depending on the drift field. The number of pulses above threshold was counted regardless of their pulse heights to enhance the S2 signal. The pulse counts are normalized to the S1 count.}
\label{fig:fig15_new}
\end{center}
\end{figure}

Figure \ref{fig:fig15_new} shows the time distribution of 
PMT pulses 
for cosmic events triggered with the PMT itself, 
obtained for four different values of the drift field, i.e. 443 V/cm (magenta), 355 V/cm (blue), 266 V/cm (green) and 178 V/cm (red). The vertical axis shows the number of PMT pulses above the threshold $\sim$0.25 p.e. regardless of their pulse heights. The pulse counts are normalized to the highest peak at 
$t = 0$ $\mu$s ($T_0$), 
which corresponds to the primary scintillation light (S1). In this way the visibility of the S2 signals can be enhanced despite the fact that they mostly consist of a bunch of single photoelectron pulses distributed over several hundreds of microseconds. 
It is worth mentioning that, to ensure a pulse linearity for a large signal, relatively large capacitances were connected to the four last dynode stages. 
A polypropylene capacitor having 22 nF was used for the dynode 9 and 10, and 122 nF for the dynode 11 and 12. 
No evidence of a gain reduction was seen after a very strong S1 pulse (see Figure \ref{fig:additional_figure}) with these capacitors.

\begin{figure}[hbtp]
\begin{center}
\includegraphics[height=20pc]{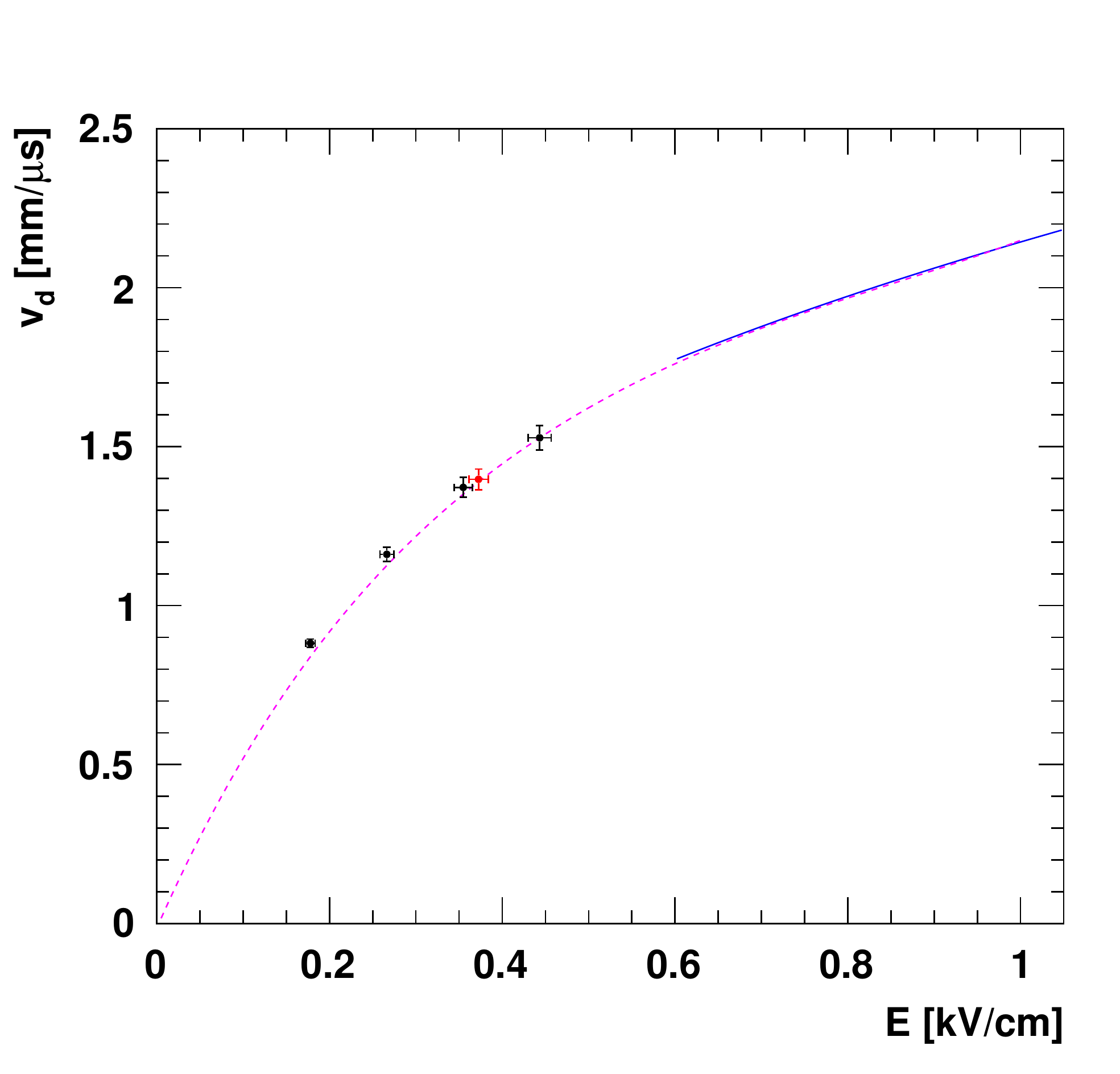}
\caption{Measured drift velocity as a function of the applied
drift field. Black points are measurements with proportional light and the red point is a measurement using the charge readout. See text.
The dotted curve is polynomial function fitted to low field measurements~\cite{Amoruso:2004dy}
rescaled to account for our operating temperature.
The solid curve shows 
the Walkowiak parametrization~\cite{Walkowiak:2000wf}  for $E\geq 0.6~kV/cm$ computed for 86.3~K.}
\label{fig:fig14}
\end{center}
\end{figure}

The S1 peak is followed by 
the S2 signal continuing until the maximum drift time which depends on the drift field, and can be seen as clear edges 
in Figure~\ref{fig:fig15_new}.
This maximum drift time $(t_{\rm max})$ corresponds to the time when the ionization electrons generated in the vicinity 
of the cathode grid, i.e. the bottom of the field cage, arrive at the LAr surface and are extracted into GAr. 
While only those muons which cross both, the top and bottom face of the field cage
create S2 signals over the full drift time ($T_0$ to $t_{\rm max}$), 
``grazing'' muons which cross one or two of the side walls produce an S2 distribution having 
a smaller width. However, the $t_{\rm max}$ edge is clearly visible when enough 
events are accumulated, 
because of the fact that there can be no extraction after $t_{\rm max}$.
Since the drift velocity $v_{\rm d}(E, T)$
is a function of electric field ($E$) and of LAr temperature ($T$), $t_{\rm max}$ also depends on those two parameters 
\begin{equation}
t_{\rm max}(E, T) - T_0 = \frac{L}{v_{\rm d}(E, T)},
\end{equation}
where $L$ is the height of the field cage, i.e. $L = 600$ mm.
The electron drift velocities $v_{\rm d}$ thus calculated from the measured $t_{\rm max}$ are plotted  with black dots in Figure~\ref{fig:fig14} 
as a function of drift electric field. 
The drift field was calculated based on the peak-to-peak value of the input AC voltage using the 
linear function obtained from the output voltage measured as a function of the input voltage, 
which is described in Section~\ref{sec:4.3}.
The horizontal error bars correspond to  $\pm3\%$
of the calculated cathode voltage 
and the vertical ones $\pm10~\mu$s in the measurement of $t_{\rm max}$.
We compare our measurements with the ones obtained with ICARUS purity monitors
at low electric fields~\cite{Amoruso:2004dy} taken at 89~K. In order to estimate 
the drift velocity at the temperature of 86.3~K of our setup obtained from the measurements using the PLC system
(see Section~\ref{sec:2.2}), we employed the 
correction $\Delta v_d \simeq -1.7\%\Delta T$~\cite{Walkowiak:2000wf}
to scale the fifth order polynomial, obtained from a fit to the
low field measurements~\cite{Amoruso:2004dy}.
Our points agree well with the polynomial function, as shown in Figure~\ref{fig:fig14}. 

\subsection{Drift velocity and uniformity based on ionizing tracks measurements}
\label{sec:4.1b}

\begin{figure}[hbtp]
\begin{center}
\begin{minipage}[t]{\columnwidth}
\begin{center}
\includegraphics[width=\columnwidth]{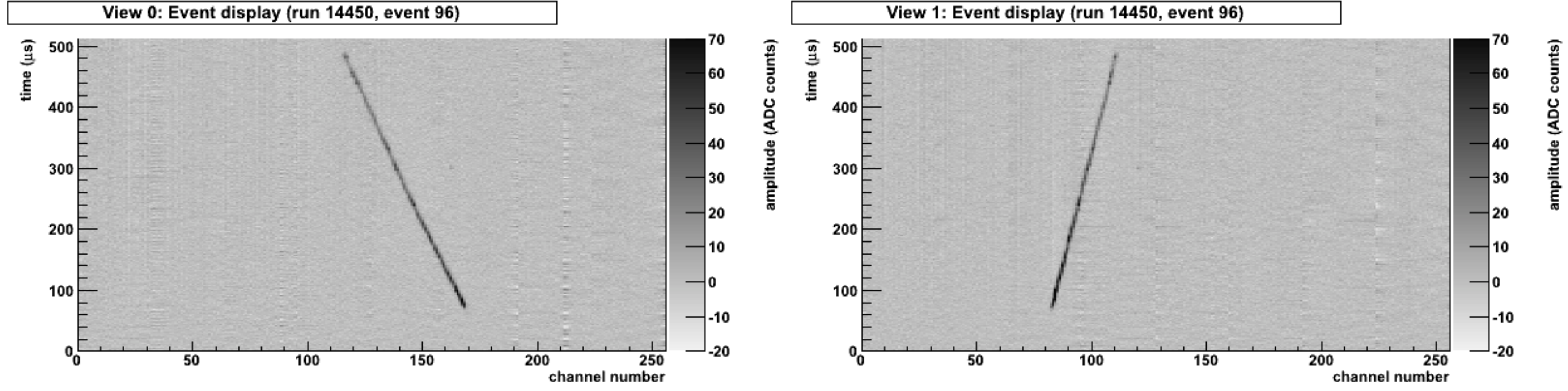}
\end{center}
\end{minipage}
\begin{minipage}[t]{\columnwidth}
\begin{center}
\includegraphics[width=\columnwidth]{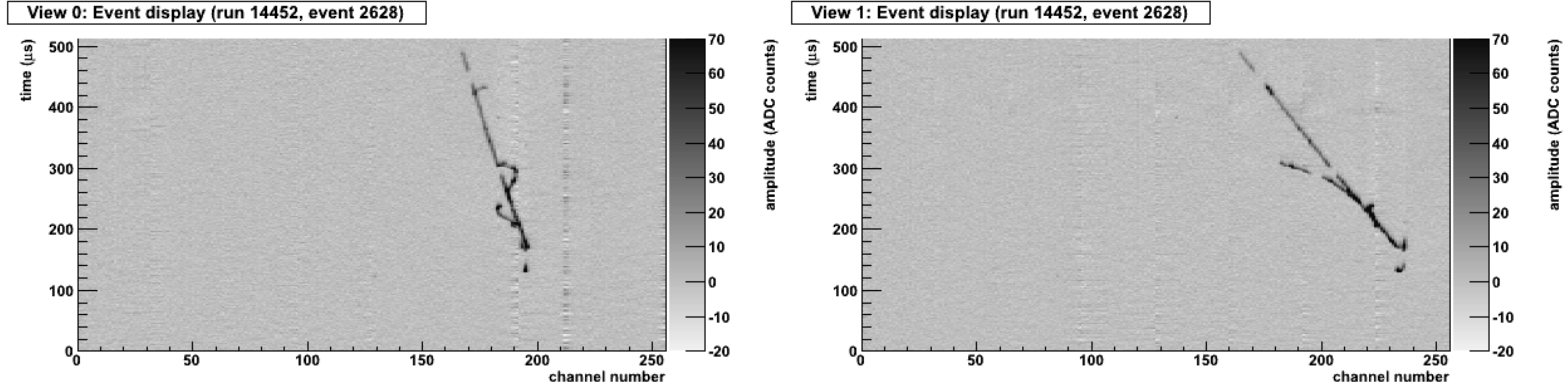}
\end{center}
\end{minipage}
\begin{minipage}[t]{\columnwidth}
\begin{center}
\includegraphics[width=\columnwidth]{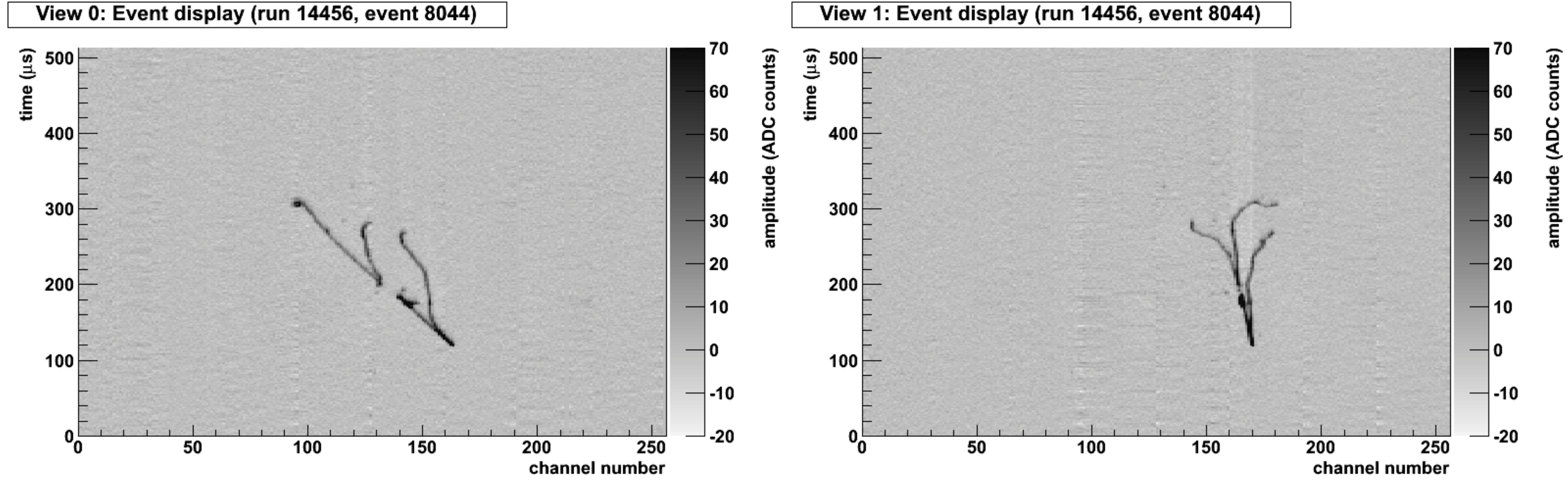}
\end{center}
\end{minipage}
\caption{Event display for three cosmic events recorded using the LEM charge readout system. 
The left and right columns correspond to the $xt$- and $yt$-projections, respectively.
Top: a straight muon track crossing the field cage from top to bottom. The time of the start ($T_0 \sim 60$ $\mu$s) and the end ($t_{\rm max} \sim 490$ $\mu$s) of the extraction are clearly visible. Middle: an example of a ``grazing'' muon track with delta rays, where the extraction starts later at $t \sim 150$ $\mu$s. 
Bottom: interacting event.}
\label{fig:fig15}
\end{center}
\end{figure}

The LEM charge readout system operated successfully in  double-phase mode. 
Three examples of cosmic events obtained with typical electric field configurations (
the drift field of 
$\approx$0.4 kV/cm 
and 35 kV/cm in the LEM holes, see Table~\ref{tab:table1}) are presented in Figure~\ref{fig:fig15}. 
Using a simple event display images for two projections, i.e. $xt$- and $yt$-projections, can be obtained for each event: $x$- and $y$-coordinates are two horizontal coordinates orthogonal to each other and $t$-coordinate corresponds to the vertical coordinate $z$. 
From a visual inspection, the effective gain of the readout is above 15 for best field configurations,
consistent with results obtained on the smaller detector prototype having an active area of 10 $\times$ 10 cm$^2$ 
\cite{Badertscher:2010zg, Badertscher:2009av}.
A quantitative analysis will be reported elsewhere~\cite{Devis:electronics}.
The maximum drift time $t_{\rm max}$ can be directly
determined from those tracks, i.e. the $t_{\rm max}$ edge is clearly visible in a time distribution of the charge signals. 
The corresponding drift velocity is shown in Figure~\ref{fig:fig14} (with a red dot)
and is in a good agreement with the other  measured points.

\begin{figure}[hbtp]
\begin{center}
\includegraphics[width=0.7\columnwidth]{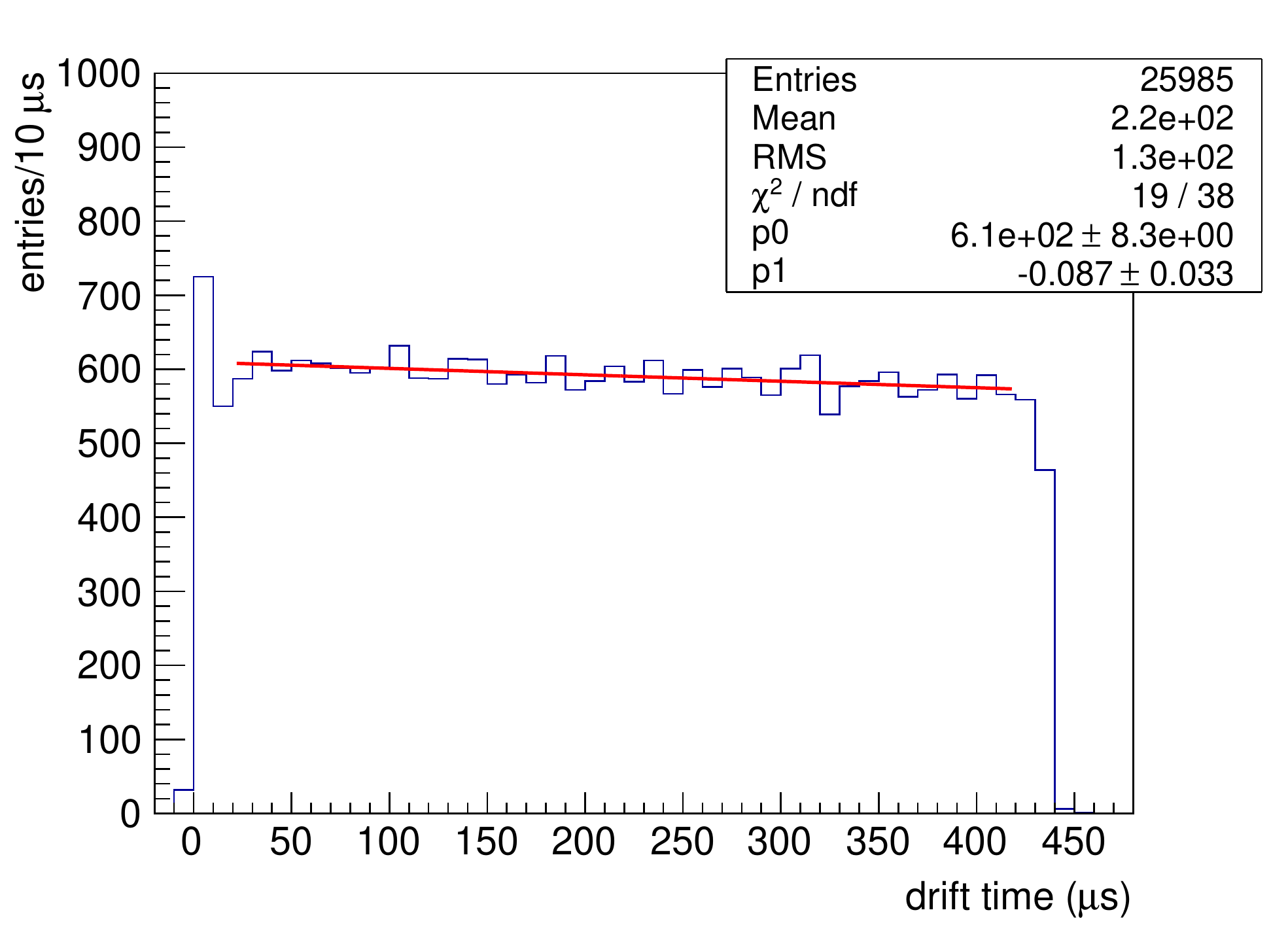}
\caption{Drift time distribution of hits recorded with the charge readout system for 147 tracks of cosmic muons. Continuous tracks crossing both the top and the bottom face of the field cage were selected. The data were taken with the typical configurations, i.e. 35 kV/cm in the LEM amplifying holes and a drift electric field of 0.4 kV/cm.
The red line shows a straight line fit (see text).}
\label{fig:fig16}
\end{center}
\end{figure}

To evaluate the uniformity of the drift electric field along the vertical coordinate,
cosmic muon tracks recorded using the LEM charge readout system were analyzed. 
%
Figure \ref{fig:fig16} shows a time distribution of the hits  for 147 fully reconstructed tracks. 
The number of readout channels giving a charge signal above threshold (i.e. a hit) within each time bin was counted and put into the histogram. 
The data were taken with the typical field configurations (c.f. Table~\ref{tab:table1}). 
The muon tracks which crossed both of the top and the bottom faces of the field cage were selected by an offline analysis. 
In addition a continuity of the track was required selecting tracks 
having no more than eight 
readout channels without a hit along the track. 

The number of hits 
in each time bin ($dN_{\rm hit}/dt$) is proportional, e.g. in the $x$-view, to $dx/dt$ and can be related to the electron drift velocity $v_{\rm d}$  
\begin{equation}
\frac{dN_{\rm hit}}{dt} \propto \frac{dx}{dt} = \frac{dx}{dz} \cdot \frac{dz}{dt} = \frac{dx}{dz} \cdot v_{\rm d} (z(t)).
\label{eq:4.2}
\end{equation}
Assuming that the track is straight, the slope $dx/dz$ 
is constant and hence $dN_{\rm hit}/dt$ is proportional to $v_{\rm d}$. 
The distribution in Figure~\ref{fig:fig16} can therefore be interpreted to be the drift velocity as a function of the vertical coordinate $z$. 
The drift velocity is constant when the drift field is uniform, i.e. any deviation from a flat distribution indicates a non-uniformity of the field. 

The excess at $t = 0$ $\mu$s is an edge effect which can be explained by the strong extraction field near the liquid argon surface. 
The rest of the distribution is straight and can be fitted with a  line of a given slope to take into account
the slight decrease towards a longer drift time, i.e. towards the bottom of the field cage. 
Fitting from 20 to 420 $\mu$s and using the average velocity $\langle v \rangle = 600/433 = 1.39$ mm/$\mu$s, one obtained 
\begin{equation}
v_{\rm d} (t) = v_0 - at, 
\end{equation}
with $v_0 = 1.43 \pm 0.02$ mm/$\mu$s and $a = (2.1 \pm 0.7)\cdot10^{-4}$ mm/$\mu$s$^2$.
The drift velocity at the bottom of the field cage thus was calculated to be $v_{\rm d} (t_{\rm max}) = 1.34 \pm 0.04$ mm/$\mu$s which is 94\%  of $v_0$. 

From the obtained drift velocity one can derive the drift electric field using our 
polynomial function shown in Figure \ref{fig:fig14} (dotted curve).
The electric field thus calculated was 
$392 \pm 10$~V/cm at the top of the field cage, respectively $350 \pm 18$ V/cm at the bottom. 
The field at the bottom was found to be ($11 \pm 5$)\% smaller with respect to that at the top. 
For the first approximation this reduction of the electric field at the bottom of the field cage can be explained by the the non-linearity of the voltage distribution to the field shapers due to the shunt capacitance in the Greinacher circuit, as described in Section \ref{sec:3.3}. 
The obtained reduction was found compatible with that of $\sim$10\% which was expected for the shunt capacitance of 4.4 pF obtained from the measurement in LAr. 
It should be noted that such a non-linearity and its influence to the field uniformity can be reduced by further optimizations of the design of the Greinacher circuit and of the field cage, respectively. It can also be corrected for offline.
The result reported above demonstrate that this method was effective in evaluating the uniformity of the drift field.
With increased statistics it 
will allow a precise correction in the conversion of the measured drift time to $z$-coordinate.

\section{Conclusions}
\label{sec:5}

We have built and operated the first ever LAr LEM-TPC having a large active area with dimensions 76 $\times$ 40 cm$^2$ and a drift length of 60~cm.
The chamber was equipped with a 30-stage immersed cryogenic Greinacher HV multiplier for generating a HV for the drift field. 
The characteristics of the HV system were extensively studied 
in air at room temperature and in LAr. 
An acceptable linearity of the output DC voltage to the input AC voltage and of the voltage distribution over the Greinacher stages was obtained from the measurements.
The LAr LEM-TPC prototype operated successfully in the double-phase (liquid-vapor) mode in pure argon during a test period of $\sim$1~month, recording 
three-dimensional images of very high visual quality from cosmic muon tracks. 
The HV system operated stably in the double-phase operation mode of the whole TPC in pure argon. 
As expected the system did not introduce noise problems on the readout electronics of the charge readout system, which is one of the advantages of this novel technique promising for a giant-scale LAr-TPC. 
The cathode voltage of $\approx$36~kV was reached leading to a drift field of $\approx$0.6 V/cm.

\acknowledgments

This work was supported by ETH Z\"{u}rich and the Swiss National Science Foundation (SNF). We are grateful to CERN for their hospitality and thank R. Oliveira and the TS/DEM group, where several of the components of our detector were manufactured. We also thank the RD51 Collaboration for useful discussions and suggestions.
Special thanks go to N. Bourgeois and the CERN PLC support group for the PLC system and to T. Schneider and Thin Film \& Glass group of CERN for helping us for the WLS coating of the PMTs. 
We thank Maria de Prado (CIEMAT), Takasumi Maruyama (KEK/IPNS), Junji Naganoma (Waseda), Hayato Okamoto (Waseda) for their participation
in the data-taking phase.


\end{document}